\documentstyle[aps,multicol,prb,epsf]{revtex}
\def\sigmav{{\mbox{\boldmath{$\sigma$}}}}
\def\rhov{{\mbox{\boldmath{$\rho$}}}}
\def\xiv{{\mbox{\boldmath{$\xi$}}}}
\def\chiv{{\mbox{\boldmath{$\chi$}}}}
\def\Bv{{\mbox{\boldmath{$B$}}}}
\def\Av{{\mbox{\boldmath{$A$}}}}
\begin{document}
\draft
\bibliographystyle{apsrev}

\title{LANDAU EXPANSION FOR THE KUGEL-KHOMSKII $t_{2g}$ HAMILTONIAN}
\author{A. B. Harris}
\address{Department of Physics and Astronomy, University of
Pennsylvania, Philadelphia, PA 19104}
\author{Amnon Aharony, O. Entin-Wohlman, and I. Ya. Korenblit}
\address{School of Physics and Astronomy, Raymond and Beverly Sackler
Faculty of Exact Sciences, \\ Tel Aviv University, Tel Aviv 69978,
Israel}
\author{Taner Yildirim}
\address{NIST Center for Neutron Research,
National Institute of Standards and Technology, Gaithersburg, MD
20899}
\date{\today}
\maketitle

\begin{abstract}
The Kugel-Khomskii (KK) Hamiltonian for the titanates describes
spin and orbital superexchange interactions between $d^1$ ions in
an ideal perovskite structure in  which the three $t_{2g}$
orbitals are degenerate in energy and electron hopping is
constrained by cubic site symmetry. In this paper we implement a
variational approach to mean-field theory in which each site, $i$,
has its own $n \times n$ single-site density matrix $\rhov(i)$,
where $n$, the number of allowed single-particle states, is 6 (3
orbital times 2 spin states). The variational free energy from
this 35 parameter density matrix is shown to exhibit the unusual
symmetries noted previously which lead to a wavevector-dependent
susceptibility for spins in $\alpha$ orbitals which is
dispersionless in the $q_\alpha$-direction.  Thus, for the cubic
KK model itself, mean-field theory does not provide wavevector
`selection',  in agreement with rigorous symmetry arguments. We
consider the effect of including various perturbations. When
spin-orbit interactions are introduced, the susceptibility has
dispersion in all directions in ${\bf q}$-space, but the resulting
antiferromagnetic mean-field state is degenerate with respect to
global rotation of the staggered spin, implying that the spin-wave
spectrum is gapless.  This possibly surprising conclusion is also
consistent with rigorous symmetry arguments. When
next-nearest-neighbor hopping is included, staggered moments of
all orbitals appear, but the sum of these moments is zero,
yielding an exotic state with long-range order without long-range
spin order. The effect of a Hund's rule coupling of sufficient
strength is to produce a state with orbital order.
\end{abstract}



\begin{multicols}{2}

\section{INTRODUCTION}

High temperature superconductivity\cite{B} and colossal
magnetoresistance\cite{CMR} have sparked much recent interest in
the magnetic properties of transition metal oxides, particularly
those with orbital degeneracy.\cite{KKrev,TN} In many transition
metal oxides, the $d$ electrons are localized due to the very
large on-site Coulomb interaction, $U$. In cubic oxide
perovskites, the crystal field of the surrounding oxygen octahedra
splits the $d$-orbitals into a two-fold degenerate $e_{g}$ and a
three-fold degenerate $t_{2g}$ manifold. In most cases, these
degeneracies are further lifted by a cooperative Jahn-Teller (JT)
distortion,\cite{KKrev} and the low energy physics is well
described by an effective superexchange spin-only model.
\cite{ANDERSON,TY1,TY2} However, some  perovskites, such as
LaTiO$_{3}$,\cite{LTO,YTO} do
not undergo a significant JT distortion, in spite of the orbital
degeneracy.\cite{noJT} In these systems, the effective
superexchange model must deal with not only the spin degrees of
freedom but also the degenerate orbital degrees of
freedom.\cite{KKrev,TN,KK} The large degeneracy of the resulting
ground states may then yield rich phase diagrams, with exotic
types of order, involving a strong interplay between the spin and
orbital sectors.\cite{TN,LTO,YTO}

In the idealized cubic model for the titanates, there is one $d$
electron in the $t_{2g}$ degenerate manifold, which contains the
wavefunctions $|X \rangle \equiv d_{yz}$, $|Y \rangle \equiv
d_{xz}$, and $|Z \rangle \equiv d_{xy}$. Following Kugel and
Khomskii (KK),\cite{KK} one starts from a Hubbard model with
on-site Coulomb energy $U$ and nearest-neighbor (nn) hopping
energy $t$. For large $U$, this model can be reduced to an
effective superexchange model, which involves only nn spin and
orbital coupling, with energies of order $\epsilon=t^2/U$. This
low energy model has been the basis for several theoretical
studies of the titanates. In particular, it has been suggested
\cite{GK1} that the KK Hamiltonian gives rise to an ordered
isotropic spin phase, and that an energy gap in the spin
excitations can be caused by spin-orbit interactions.\cite{GK2}
However, these papers are based on assumptions and approximations
which are hard to assess. Recently\cite{PRL} (this will be
referred to as I) we have presented rigorous symmetry arguments
which show several unusual symmetries of the cubic KK Hamiltonian.
Perhaps the most striking symmetry is the rotational invariance of
the total spin of $\alpha$ orbitals (where $\alpha = X,Y$, or $Z$)
summed over all sites in a plane perpendicular to the
$\alpha$-axis.  This symmetry implies that in the disordered phase
the wavevector-dependent spin susceptibility for $\alpha$
orbitals, $\chi_\alpha({\bf q})$ is dispersionless in the
$q_\alpha$-direction. In addition, as discussed in I, this
symmetry implies that the system does not support long-range spin
order at any nonzero temperature. Thus the idealized cubic KK
model is an inappropriate starting point to describe the
properties of existing titanate systems.  This peculiar rotational
invariance
depends on the special symmetry of the hopping matrix element and
it can be broken by almost any perturbation such as rotation of
the oxygen octahedra.  Here we consider the effect of
symmetry-breaking perturbations due to a) spin-orbit interactions,
b) next-nearest-neighbor (nnn) hopping, and c) Hund's rule
coupling. According to the general symmetry argument of I,
although long-range order at nonzero temperature is possible when
spin-orbit interactions are included, the system still possesses
enough rotation symmetry that the excitation spectrum should be
gapless.  (This conclusion is perhaps surprising because once
spin-orbit interactions are included, the system might be expected
to distinguish directions relative to those defined by the
lattice.)  This argument would imply that mean-field theory will
produce a state which has a continuous degeneracy associated with
global rotation of the spins. The purpose of this paper is to
implement mean-field theory and to interpret the results obtained
therefrom in light of the general symmetry arguments. We will
carry out this analysis using the variational properties of the
density matrix. In a separate paper\cite{III}  (which we will
refer to as III, the present paper being paper II) we will study
the self-consistent equations of mean-field theory which contain
information equivalent to what we obtain here, but in a form which
is better suited to a study of the ordered phase.   Here our
analysis is carried out for the cubic KK Hamiltonian with and
without the inclusion of the symmetry-breaking perturbations
mentioned above. In the presence of spin-orbit interactions we
find that the staggered moments of different orbital states are
not collinear, so that the net spin moment is greatly reduced from
its spin-only value. The effect of  nnn hopping is also
interesting. Within mean-field theory, this perturbation was found
to stabilize a state having long-range staggered spin order for
each orbital state, but the staggered spins of the three orbital
states add to zero. When only Hund's rule coupling is included,
mean-field theory predicts stabilization of long-range spin and
orbital order.  However, elsewhere\cite{IV} we
show that fluctuations favor spin-only order.  As a result, a
state with long-range order of both spin and orbital degrees of
freedom can only occur when the strength of the Hund's rule
coupling exceeds some critical value which we can not estimate in
the present formalism.

Briefly this paper is organized as follows.  In Sec. II we discuss
the KK Hamiltonian and fix the notation we will use. In Sec. III
we discuss the construction of the mean-field trial density matrix
as the product of single-site density matrices, each of which acts
on the space of six one-electron states of an ion, and whose
parametrization therefore requires 35 parameters. Here we show
that the wavevector-dependent spin susceptibilities which diverge
as the temperature is lowered through a critical value have
dispersionless directions, so that unusually mean-field theory
provides no `wavevector selection' at the mean-field transition.
In Sec. IV we discuss the Landau expansion at quartic order. In
Sec. V we treat several lower symmetry perturbations, namely
spin-orbit interactions, nnn hopping, and Hund's rule coupling. In
each of these cases `wavevector selection' leads to the usual
two-sublattice structure,  but the qualitative nature of ordering
depends on which perturbation is considered.  In Sec. VI we
summarize our work and discuss its implications.

\section{THE HAMILTONIAN}

\begin{figure}
\leavevmode \epsfclipon \epsfxsize=5.8truecm
\vbox{\epsfbox{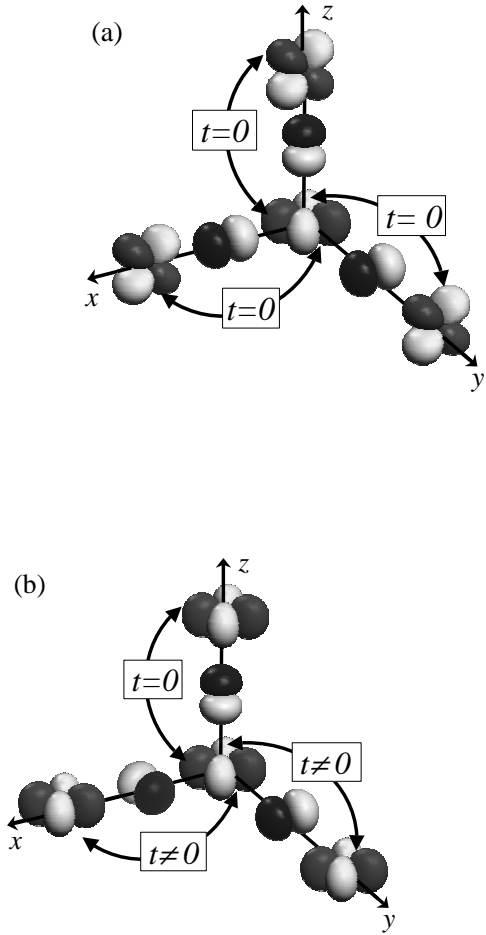}}
\caption{A schematic view of the $|Z\rangle = d_{xy}$ orbitals and
the (indirect) hopping parameter $t$ via intermediate oxygen
p-orbitals. Positive (negative) regions of wavefunctions are
represented by dark (light) lobes. 
In (a) we show that the hopping matrix elements
between orbitals of different flavors are zero.
In (b) we show that
there is no indirect hopping along the z-axis for an electron 
in the Z-orbital, due to
symmetry. 
} 
\label{DXY}
\end{figure}

The system we treat is a simple cubic lattice of ions with one d
electron per ion in a d-band whose five orbital states are split
into an $e_{ g}$ doublet at high energy and a $t_{2g}$ triplet at
low energy.  Following the seminal work of Kugel and
Khomskii\cite{KK} (KK), we describe this system by a Hubbard
Hamiltonian ${\cal H}_H$ of the form
\begin{eqnarray}
{\cal H}_H &=& \sum_{i \alpha \sigma} \epsilon_\alpha c_{i \alpha
\sigma}^\dagger c_{i \alpha \sigma} + \sum_{\langle ij \rangle}
\sum_{\alpha \beta \sigma} t_{\alpha \beta} (i,j) c_{i \alpha
\sigma}^\dagger c_{j \beta \sigma} \nonumber \\ && \ + U \sum_i
\sum_{\alpha \leq \beta} \sum_{\sigma \sigma'} c_{i \alpha \sigma
}^\dagger c_{i \alpha \sigma} c_{i \beta \sigma'}^\dagger c_{i
\beta \sigma'} \ , \label{HHUB}
\end{eqnarray}
where $c_{i \alpha \sigma}^\dagger$ creates an electron in the
orbital labeled $\alpha$ in spin state $\sigma$ on site $i$,
$\epsilon_\alpha$ is the crystal field energy of the $\alpha$
orbital, $t_{\alpha \beta}(i,j)$ is the matrix element for hopping
between orbital $\alpha$ of site $i$ and orbital $\beta$ of site
$j$, and $\langle ij\rangle$ indicates that the sum is over pairs
of nearest neighboring sites $i$ and $j$ on a simple cubic
lattice.  It is convenient to refer to the orbital state of an
electron as its `flavor'. In this terminology $c_{i \alpha
\sigma}^\dagger$ creates an electron of flavor $\alpha$ and
$z$-component of spin $\sigma$ on site $i$.  Initially we consider
the case when the Coulomb interaction does not depend on which
orbitals the electrons are in. In a later section we will consider
the effects of Hund's-rule coupling. In a cubic crystal field, the
crystal-field energy $\epsilon_\alpha$ splits the five orbital d
states into a low-energy triplet, whose states are $d_{yz}\equiv
X$, $d_{xz}\equiv Y$, and $d_{xy}\equiv Z$, and a high energy
doublet, whose presence is ignored.  In this model it is assumed
that hopping occurs only between nearest neighbors and proceeds
via superexchange through an intervening oxygen $p$ orbital, so
that the symmetry of the hopping matrix is that illustrated in
Fig. \ref{DXY}. Thus $t_{\alpha \beta}$ is zero if $\alpha
\not=\beta$ and $t_{\alpha \alpha}(i,j)=t$, except that $t_{\alpha
\alpha}(i,j)$ vanishes if  the bond $\langle ij \rangle$ is
parallel to the $\alpha$-axis.\cite{KK} The $\alpha$-axis is
called\cite{HUND} the inactive axis for hopping between $\alpha$
orbitals.  When $t \ll U$, KK reduced the above Hubbard
Hamiltonian to an effective Hamiltonian for the manifold of states
for which each site has one electron in a $t_{ 2g}$ orbital state.
We will call this low-energy Hamiltonian the KK Hamiltonian and it
can be regarded as a many-band generalization of the Heisenberg
Hamiltonian.  The KK Hamiltonian is often written in terms of spin
variables to make the analogy with the Heisenberg model more
apparent, but for our purposes it is more convenient to write the
(KK) Hamiltonian in the form
\begin{eqnarray}
{\cal H}_{KK} &=& \epsilon \sum_{\langle ij \rangle} \sum_{\beta
\gamma \not= \langle ij \rangle} \sum_{\eta \rho} c_{i \beta
\eta}^\dagger c_{i \gamma \rho}
c_{j \gamma \rho}^\dagger c_{j \beta \eta} \nonumber \\
&\equiv& \ \epsilon \sum_{\langle ij \rangle}
\sum_{\beta \gamma \not= \langle ij \rangle} \sum_{\eta \rho}
Q_{\beta \eta;  \gamma \rho}(i) Q_{\gamma \rho ;  \beta \eta }(j) \ ,
\end{eqnarray}
where $\epsilon = t^2 /U$ and the notation $\beta \gamma \not= \langle ij \rangle$
indicates that in the sum over $\beta$ and $\gamma$ neither of these
are allowed to be the same as the coordinate direction of the bond
$\langle ij \rangle$.

\vspace{1cm}

\leavevmode \epsfclipon \epsfxsize=6.truecm
\hspace{1cm}\vbox{\epsfbox{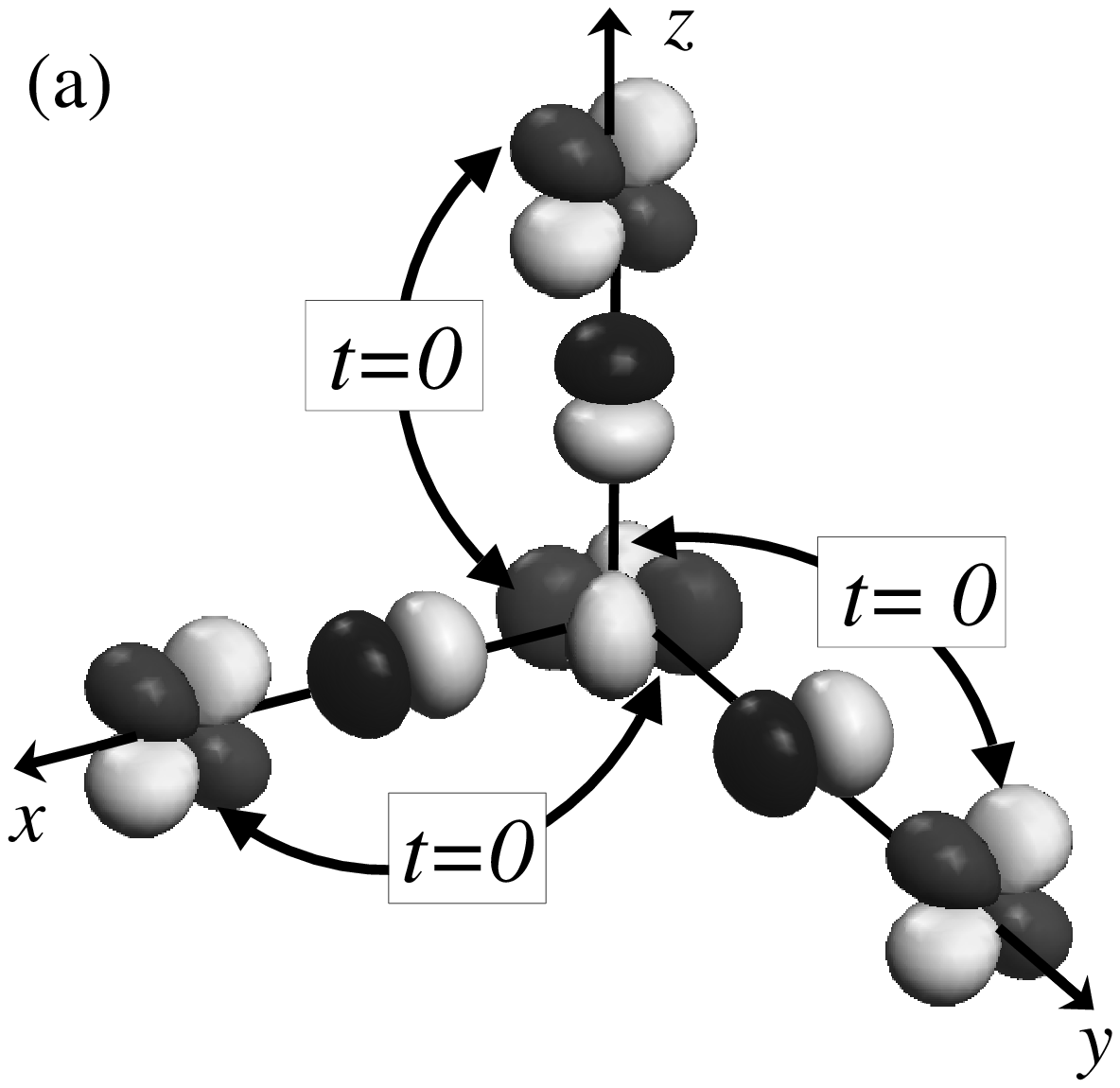}}

\vspace{1cm}


Previously\cite{PRL} we pointed out several unusual symmetries of
this Hamiltonian.  By an $\alpha$-plane we mean any plane
perpendicular to the $\alpha$ axis (which is the inactive axis for
$\alpha$-hopping).  In I we showed that the total number of
electrons in an $\alpha$-plane which are in $\alpha$ orbitals is
constant.  In addition, the total spin vector (as well as its $z$
component) summed over all electrons in $\alpha$ orbitals in any
given $\alpha$-plane was  shown to be a good quantum number. The
fact that one can rotate the spin of all $\alpha$ electrons (these
are electrons in $\alpha$ orbitals) in any $\alpha$-plane at no
cost in energy implies that there is no long-range spin order at
any nonzero temperature.\cite{PRL} Nevertheless, since
experiment\cite{LTO} shows that LaTiO$_3$ does exhibit long-range
spin order, it must be that spin ordering is caused by some,
possibly small, symmetry breaking perturbation, which should be
added to the idealized KK model. Therefore it is worthwhile
investigating what form of long-range order results when possible
symmetry-breaking perturbations are included.  Although the
mean-field results we obtain below should not be taken
quantitatively, they may form a qualitative guide to the type of
ordering one might expect for more realistic extensions of the
above KK model. We also noted\cite{TY1,PRL} that even when
spin-orbit coupling is included, the Hamiltonian has sufficient
symmetry that the spin-wave spectrum remains gapless. As a result,
the gap observed\cite{LTO} in the excitation spectrum of LaTiO$_3$
can not be explained on the basis of the KK Hamiltonian with only
the spin-orbit interaction as a perturbation. As we shall see,
these symmetries are realized by the mean-field solutions we
obtain.

\section{LANDAU EXPANSION AT QUADRATIC ORDER}

We will develop the Landau expansion of the free energy as a
multivariable expansion in powers of the full set of order
parameters necessary to describe the free energy arising from the
KK Hamiltonian.  In this section we construct this expansion up to
quadratic order in these order parameters and thereby analyze the
instability of the disordered phase relative to arbitrary types of
long-range order.  In later sections we discuss how this picture
is modified by higher-order terms in the expansion,  and by the
addition of various symmetry-breaking terms into the Hamiltonian.

\subsection{Parametrizing the Density Matrix}

The version of mean-field theory which we will implement is
based on the variational principle according to which the
exact free energy is obtained by minimizing the free energy
functional $F(\rhov)$ as a function of the trial density matrix
$\rhov$, which must be Hermitian, have no negative eigenvalues,
and be normalized by ${\rm Tr} \rhov =1$.  Here the trial free energy is
\begin{eqnarray}
F(\rhov) &=& {\rm Tr} \Bigl[ \rhov \bigl( {\cal H} + kT \ln \rhov
\bigr) \Bigr]  ,
\end{eqnarray}
where the first term is the trial energy and the second is $-T$
times the trial entropy, where $T$ is the temperature.
Mean-field theory is obtained by the ansatz that $\rhov$ is the
product of single-site density matrices, $\rhov(i)$:
\begin{eqnarray}
\rhov = \prod_i \rhov(i)  ,
\end{eqnarray}
and $F(\rhov)$ is then minimized with respect to the variables
used to parametrize the density matrix, $\rhov (i)$. Since $\rhov
(i)$ acts in the space of $t_{2g}$ states of one electron, it is a
$6 \times 6$ dimensional Hermitian matrix with unit trace.

The most general trial density matrix (for site $i$) can be written
in the form
\begin{eqnarray}
 \rhov(i)=\frac{1}{6}{\cal I}+X(i),\label{rho}
\end{eqnarray}
where
\begin{eqnarray}
X(i)=\sum_{\alpha \beta}\sum_{\rho
\eta}c^{\dagger}_{i\alpha\rho}Y_{\alpha\rho\beta\eta}(i)c_{i\beta\eta}\
,\label{par}
\end{eqnarray}
with
\begin{eqnarray}
Y_{\alpha\rho\beta\eta}(i)={\bf A}_{\alpha\beta}(i)\delta_{\rho
\eta}+\vec{\bf B}_{\alpha\beta}(i)\cdot\vec{\sigmav}_{\rho \eta}\
.\label{Y}
\end{eqnarray}
Here $\vec{\sigmav}$ is  the Pauli matrix vector, and ${\bf
A}_{\alpha\beta}(i)$, ${\bf B}^{x}_{\alpha\beta}(i)$, ${\bf
B}^{y}_{\alpha\beta}(i)$, and ${\bf B}^{z}_{\alpha\beta}(i)$ are
$3\times 3$ Hermitian matrices, of which the first is traceless.
The diagonal terms of the matrix ${\bf A}$ are parametrized for
later convenience as
\begin{eqnarray}
{\bf
A}_{xx}(i)=\frac{a_{1}(i)}{\sqrt{6}}&+&\frac{a_{2}(i)}{\sqrt{2}},\
\
{\bf A}_{yy}(i)=\frac{a_{1}(i)}{\sqrt{6}}-\frac{a_{2}(i)}{\sqrt{2}},\nonumber\\
{\bf A}_{zz}(i)&=&-\Av_{xx}(i)-\Av_{yy}(i),\label{map}
 \end{eqnarray}
such that
\begin{eqnarray}
\Av_{xx}^{2}(i)+\Av_{yy}^{2}(i)+\Av_{xzz}^{2}(i)&=&a_{1}^{2}(i)+a_{2}^{2}(i),\nonumber\\
-\Av_{xx}^{2}(i)-\Av_{yy}^{2}(i)+2\Av_{zz}^{2}(i)&=&a_{1}^{2}(i)-a_{2}^{2}(i).
\end{eqnarray}

For any operator ${\cal O}(i)$ associated with site $i$
we define
\begin{eqnarray}
\langle {\cal O}(i) \rangle \equiv {\rm Tr} [ {\cal O}(i) \rhov ]
\ ,
\end{eqnarray}
where Tr denotes a trace over the six states $|\alpha , \sigma
\rangle$ of the atom at site $i$ with a single $t_{ 2g}$ electron.
Then the diagonal matrix elements of ${\bf A}(i)$ give the
occupations of orbital states,
\begin{eqnarray}
\langle N_\alpha (i)\rangle =\langle\sum_\sigma c_{i\alpha
\sigma}^\dagger c_{i \alpha \sigma}\rangle =2\Av_{\alpha\alpha
}(i),\label{N}
\end{eqnarray}
which may be related to the matrix elements of the angular
momentum, $L$,
\begin{eqnarray}
\langle \frac{L_{x}^{2}(i)-1}{3}\rangle =\langle N_x(i) \rangle
&=& {1 \over 3} + {2 \over \sqrt 6} a_1(i)
+ \sqrt 2 a_2(i),\nonumber\\
\langle \frac{L_{y}^{2}(i)-1}{3}\rangle = \langle N_y(i) \rangle
&=& {1 \over 3} + {2 \over \sqrt 6} a_1(i)
- \sqrt 2 a_2(i),\nonumber\\
\langle \frac{L_{z}^{2}(i)-1}{3}\rangle = \langle N_z(i) \rangle
&=& {1 \over 3} - {4 \over \sqrt 6} a_1(i) \ .\label{NEQ}
\end{eqnarray}
The off-diagonal matrix elements of  ${\bf A}(i)$ are
\begin{eqnarray}
\langle L_{\gamma}(i)\rangle &=&i \sum_{\alpha\beta} \sum_{\sigma}\langle
  c_{i\alpha\sigma}^{\dagger}c_{i\beta\sigma}
\rangle \epsilon_{\alpha\beta\gamma}
\nonumber\\
&=&-2i\sum_{\alpha\beta}\Av_{\alpha\beta}(i)\epsilon_{\alpha\beta\gamma},
\end{eqnarray}
where $\epsilon_{\alpha\beta\gamma}$ is the fully antisymmetric
tensor.
Similarly,
\begin{eqnarray}
\langle L_{\beta}(i)L_{\gamma}(i)+L_{\gamma}(i)L_{\beta}(i)\rangle
&=&-3\sum_{\sigma}\langle
c_{i\beta\sigma}^{\dagger}c_{i\gamma\sigma}+c^{\dagger}_{i\gamma\sigma}c_{i\beta\sigma}\rangle\nonumber\\
&=&-6[\Av_{\beta\gamma}(i)+\Av_{\gamma\beta}(i)].
\end{eqnarray}
Similarly, the diagonal matrix elements of ${\bf B}^{\gamma}(i)$,
${\bf B}^{\gamma}_{\alpha\alpha}(i)$, give the thermal expectation
value of the $\gamma$ component of the spin of $\alpha$-flavor
electrons:
\begin{eqnarray}
\langle S_{\alpha \gamma}(i)\rangle &=&\sum_{\sigma\eta} \langle
c_{i \alpha \sigma}^\dagger \sigmav_{\sigma \eta}^\gamma c_{i
\alpha \eta} \rangle \ =2{\bf
B}^{\gamma}_{\alpha\alpha}(i).\label{S}
\end{eqnarray}
The off-diagonal matrix elements of ${\bf B}^{\gamma}(i)$ are
related to the order-parameters associated with correlated
ordering of spins and orbits.

 In general, the density matrix Eq.
(\ref{rho}) yields the average
\begin{eqnarray}
&&\langle Q_{\alpha \sigma ; \beta \eta} (i) \rangle \equiv
\langle c_{i \alpha \sigma}^\dagger c_{i\beta \eta}\rangle
=\frac{1}{6}\delta_{\alpha\beta}\delta_{\sigma\eta}\nonumber\\
&+&\sum_{\stackrel{\alpha '\beta '}{\rho \tau}}\langle
c^{\dagger}_{i\alpha\sigma}c_{i\beta\eta}c^{\dagger}_{i\alpha
'\rho}  ({\bf A}_{\alpha '\beta '}(i)\delta_{\rho \tau}+\vec{\bf
B}_{\alpha '\beta '}(i)\cdot\vec{\sigmav}_{\rho \tau} )c_{i\beta
'\tau}\rangle\nonumber\\
&=&\delta_{\alpha\beta}\delta_{\sigma\eta}/6+{\bf A}_{\beta\alpha}
(i)\delta_{\sigma\eta}+\vec{\bf
B}_{\beta\alpha}(i)\cdot\vec{\sigmav}_{\eta\sigma}.\label{AVERAGE}
\end{eqnarray}

\subsection{Construction of the Trial Free Energy}

Using the result Eq. (\ref{AVERAGE}), we get the trial energy,
$U$, as
\begin{eqnarray}
&&U = \ \epsilon \sum_{\langle ij \rangle } \sum_{\beta \gamma
\not= \langle ij \rangle} \sum_{\eta \rho} \langle Q_{\beta \eta;
\gamma \rho}(i)\rangle \langle Q_{\gamma \rho ;  \beta \eta }(j)
\rangle
\nonumber\\
&=&2\epsilon\sum_{\langle ij\rangle}\sum_{\alpha\beta\neq\langle
ij\rangle}
[\Av_{\alpha\beta}(i)\Av_{\beta\alpha}(j)+\vec{\Bv}_{\alpha\beta}(i)\cdot\vec{\Bv}_{\beta\alpha}(j)
],
\end{eqnarray}
where we have used the identity
\begin{eqnarray}
&&\sum_{\rho_{2}}(\vec{\Bv}_{\alpha_{1}\alpha_{2}}
\cdot\vec{\sigmav}_{\rho_{1}\rho_{2}})(\vec{\Bv}_{\alpha_{2}\alpha_{3}}\cdot
\vec{\sigmav}_{\rho_{2}\rho_{3}})\nonumber\\
&=&\vec{\Bv}_{\alpha_{1}\alpha_{2}}\cdot\vec{\Bv}_{\alpha_{2}\alpha_{3}}\delta_{\rho_{1}\rho_{3}}
+i\vec{\sigma}_{\rho_{1}\rho_{3}}\cdot\vec{\Bv}_{\alpha_{1}\alpha_{2}}\times\vec{\Bv}_{\alpha_{2}\alpha_{3}}.
\label{identity}
\end{eqnarray}
Here and below we drop terms independent of the trial
order-parameters.

Using Eq. (\ref{rho}) we write the trial entropy as
\begin{eqnarray}
-TS &=& kT \sum_i {\rm Tr} ( 3 X^{2}(i)  - 6 X^{3}(i) + 18
X^{4}(i) + \dots ), \label{XEQ}
\end{eqnarray}
where we noted that ${\rm Tr} X(i) =0$.   The second-order
contribution is found from
\begin{eqnarray}
&&{\rm Tr} [X^{2}(i)]
=\sum_{\stackrel{\alpha\beta}{\alpha '\beta
'}}\sum_{\stackrel{\eta\rho }{\eta '\rho '}}{\rm Tr}[
c^{\dagger}_{i\alpha\rho}
Y_{\alpha\rho\beta\eta}(i)c_{i\beta\eta}\nonumber\\
&&\times c^{\dagger}_{i\alpha '\rho '}Y_{\alpha '\rho '\beta '\eta
' }(i)c_{i\beta '\eta
'}]
=\sum_{\alpha\beta}\sum_{\eta\rho
}Y_{\alpha\rho\beta\eta}(i)Y_{\beta\eta\alpha\rho}(i)\nonumber\\
&=&\sum_{\alpha\beta} [\Av_{\alpha\beta}(i)\Av_{\beta\alpha}(i)
+\vec{\Bv}_{\alpha\beta}(i)\cdot\vec{\Bv}_{\beta\alpha}(i) ].
\end{eqnarray}

At quadratic order the trial free-energy, $F=F_2$, is thus
\begin{eqnarray}
F_{2}&=&\frac{1}{2}\sum_{ij}\sum_{\alpha\beta}\chi^{-1}_{\alpha\beta}(i,j)
[\Av_{\alpha\beta}(i)\Av_{\beta\alpha}(j)
\nonumber\\
&&\ \ \ \ \ \ \ \ \ \
+\vec{\Bv}_{\alpha\beta}(i)\cdot\vec{\Bv}_{\beta\alpha}(j)],
\end{eqnarray}
where the inverse susceptibility is given by
\begin{eqnarray}
\chi_{\alpha\beta}^{-1} (i,j)
 =12kT\delta_{ij}+2\epsilon\gamma_{ij}(1-\delta_{ij,\alpha})(1-\delta_{ij,\beta}).
\end{eqnarray}
Here $\gamma_{ij}$ is unity if sites $i$ and $j$ are nearest
neighbors and  is zero otherwise, and $\delta_{ij,\alpha}$ is
unity if the bond $\langle ij\rangle$ is along the
$\alpha$-direction and is zero otherwise.

\subsection{Stability Analysis - Wavevector Selection}

We now carry out a stability analysis of the disordered phase.  At
quadratic order in the Landau expansion, possible phase
transitions from the disordered phase to a phase with long-range
order are signalled by the divergence of a susceptibility.
Depending on the higher-than-quadratic order terms in the Landau
expansion, such a transition may (or may not) be preempted by a
first-order (discontinuous) phase transition. So mean-field theory
is a simple and usually effective way to predict the nature of the
ordered phase in systems where it may not be easy to guess it.
To implement the stability analysis
we diagonalize the inverse susceptibility matrix by going to
Fourier transformed variables, whose generic definition is
\begin{eqnarray}
F({\bf q}) &=& {1 \over \sqrt N} \sum_i F({\bf r}_i) e^{-i {\bf q}
\cdot {\bf r}_i}  ,\ \
\nonumber \\
F({\bf r}_i) &=& {1 \over \sqrt N} \sum_{\bf q} F({\bf q}) e^{i
{\bf q} \cdot {\bf r}_i}  , \label{FOURIER}
\end{eqnarray}
where
$N$ is the total number of lattice sites.  Then the free energy at
quadratic order
is $F_2=\sum_{\bf q}F_2({\bf q})$, where
\begin{eqnarray}
F_{2}&=&\frac{1}{2}\sum_{\bf
q}\sum_{\alpha\beta}\chi^{-1}_{\alpha\beta}({\bf q})
[\Av_{\alpha\beta}({\bf q})\Av_{\beta\alpha}(-{\bf
q})\nonumber\\
&&\ \ \ \ \ \ \ \ \ \ \ +\vec{\Bv}_{\alpha\beta}({\bf
q})\cdot\vec{\Bv}_{\beta\alpha}(-{\bf q})],\label{FEQ}
\end{eqnarray}
with
\begin{eqnarray}
&&\chi^{-1}_{\alpha\beta}({\bf q})=12kT
\nonumber\\
&+&2\epsilon\sum_{{\bf R}_{nn}} e^{-i{\bf q}\cdot{\bf
R}_{nn}}(1-\delta_{{\bf R}_{nn},a\hat{\alpha}})(1-\delta_{{\bf
R}_{nn},a\hat{\beta}}),
\end{eqnarray}
where ${\bf R}_{nn}$ is a  vector to a nearest-neighbor site,
and $\hat{\alpha}$ is the unit vector in the $\alpha$-direction.
We hence see that we have only two kinds of inverse
susceptibilities, the one for the diagonal elements, namely
\begin{eqnarray}
\chi^{-1}_{\alpha\alpha}({\bf q})&=&12 kT+2\epsilon\sum_{{\bf
R}_{nn}} e^{-i{\bf q}\cdot{\bf R}_{nn}}(1-\delta_{{\bf
R}_{nn},a\hat{\alpha}})\nonumber\\
&=&12kT+2\epsilon
\sum_{\beta\gamma}\epsilon_{\alpha\beta\gamma}^{2}(c_{\beta}+c_{\gamma}),
\label{chid}
\end{eqnarray}
and the second for the off-diagonal matrix elements, namely
\begin{eqnarray}
\chi^{-1}_{\alpha\beta}({\bf q})=12 kT&+&2\epsilon\sum_{{\bf
R}_{nn}} e^{-i{\bf q}\cdot{\bf R}_{nn}}
(1-\delta_{{\bf R}_{nn},a\hat{\alpha}}-\delta_{{\bf
R}_{nn},a\hat{\beta}})\nonumber\\
&=&12kT+4\epsilon\sum_{\gamma}\epsilon_{\alpha\beta\gamma}^{2}c_{\gamma},
\label{chind}
\end{eqnarray}
where $c_{\alpha}\equiv \cos (q_{\alpha }a)$.

\begin{figure}
\leavevmode \epsfclipon \epsfxsize=6.truecm
\hspace{1cm}\vbox{\epsfbox{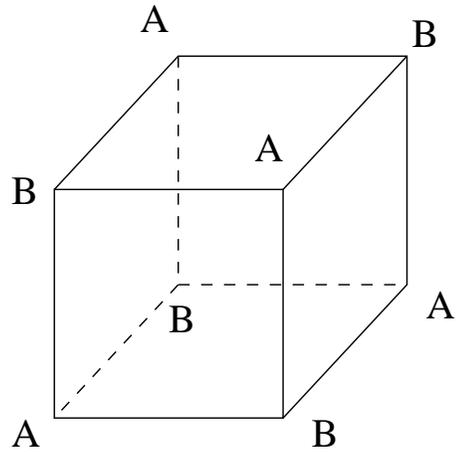}} \vspace{.5cm} \caption{The two
sublattice ``G'' state which consists of two interpenetrating
simple cubic lattices on each site of which the ions are in a
given state, either A or B.} \label{G}
\end{figure}

At high temperature all the eigenvalues of the susceptibility
matrix are finite and positive.  As the temperature is reduced,
one or more eigenvalues may become zero, corresponding to an
infinite susceptibility.  Usually this instability will occur at
some value of wavevector (or more precisely at the star of some
wavevector), and this set of wavevectors describes the periodicity
of the ordered phase near the ordering transition. This phenomenon
is referred to as `wavevector selection'.  In addition, and we
will later see several examples of this, the eigenvector
associated with the divergent susceptibility contains information
on the qualitative nature of the ordering.  Here, a central
question which the eigenvector addresses, is whether the ordering
is in the spin sector, the orbital sector, or both sectors.
If the unstable eigenvector is degenerate, one can usually
determine the symmetries  which give rise to Goldstone (gapless)
excitations. (We will meet this situation in connection with our
treatment of spin-orbit interactions.)  In the present case, we
see from Eqs. (\ref{chid}) and (\ref{chind}) that the
instabilities (where an inverse susceptibility vanishes) first
appear at $kT=2\epsilon/3$ for the diagonal susceptibilities.
Consider first the susceptibilities for unequal occupations of the
three orbital states. Making use of Eqs. (\ref{map}) and
(\ref{chid}), we write
\begin{eqnarray}
&&\sum_{\alpha}\chi^{-1}_{\alpha\alpha}({\bf
q})\Av_{\alpha\alpha}({\bf q})\Av_{\alpha\alpha}(-{\bf
q})\nonumber\\
&=&\left [\begin{array}{cc}a_{1}({\bf q})&a_{2}({\bf
q})\end{array}\right ]\chiv_{n}({\bf q})
\left [\begin{array}{c}a_{1}(-{\bf q})\\a_{2}(-{\bf
q})\end{array}\right ],
\end{eqnarray}
with the 2$\times$2 susceptibility matrix $\chiv_{n}$ given by
\begin{eqnarray}
\chi^{-1}_{n,11}({\bf q})&=&12kT+\frac{2\epsilon}{3}\Bigl (5
c_{x}+5
c_{y}+2 c_{z}\Bigr ),\nonumber\\
\chi^{-1}_{n,22}({\bf q})&=&12kT+2\epsilon\Bigl ( c_{x}+
c_{y}+2 c_{z}\Bigr ),\nonumber\\
\chi^{-1}_{n,12}({\bf q})&=&\chi^{-1}_{n,21}({\bf q})=
\frac{2\epsilon}{\sqrt{3}}\Bigl (c_{y}-c_{x}\Bigr ).
\end{eqnarray}
The instability occurs for both eigenvalues of the inverse
susceptibility matrix $\chi^{-1}_{n,\ell m}({\bf q})$, 
but only when the wavevector ${\bf q}$ assumes its
antiferromagnetic value ${\bf Q} = (\pi, \pi, \pi)/a$ which leads
to a two sub-lattice structure (see Fig. \ref{G}) called the ``G''
state. The two-fold degeneracy is the symmetry associated with
rotations in occupation number space $\langle N_x\rangle$,
$\langle N_y \rangle$, and $\langle N_z \rangle$ with the
constraint that the sum of these occupation numbers is unity. (At
quadratic order we do not yet feel the discrete cubic symmetry of
the orbital states.) In contrast, the inverse spin susceptibility
$\chi^{-1}_{\alpha\alpha}$ of Eq. (\ref{chid}) has a flat branch
so that it
vanishes for $kT=2\epsilon/3$ for any value of $q_\alpha$, when
the two other components of ${\bf q}$ assume the antiferromagnetic
value $\pi/a$.  This wavevector dependence indicates that
correlations in the spin susceptibility
become long ranged in an $\alpha$-plane, but different
$\alpha$-planes are completely uncorrelated.  Note that beyond the
fact that there is no wavevector selection in the spin
susceptibility, one has complete rotational invariance in
$\Bv^{\gamma}_{\alpha \alpha}({\bf q})$ for the components labeled
by $\gamma$ {\it independently for each orbital labeled} $\alpha$.
This result reflects the exact symmetry of the Hamiltonian with
respect to rotation of the total spin in the $\alpha$ orbital
summed over all spins in any single $\alpha$-plane.\cite{PRL} 
If we restrict attention to the G wavevector ${\bf q}={\bf Q}$, we
have complete rotational degeneracy in the 11 dimensional space
consisting of the nine $\Bv^{\gamma}_{\alpha \alpha}({\bf Q})$
spin order-parameters
and the two $a_n({\bf Q})$ occupational order-parameters.  Thus at
this level of approximation, we have O(11) symmetry!  Most of
this symmetry only holds at quadratic order in mean-field theory.
As usual, we expect that fourth (and higher) order terms in the
Landau expansion will generate anisotropies in this 11-dimensional
space to lower the symmetry to the actual cubic symmetry of the
system. As we will see, the anisotropy which inhibits the mixing
of spin and orbit degrees of freedom is not generated by the
quartic terms in the free energy. Perhaps unexpectedly, as we show
elsewhere,\cite{IV} this anisotropy is only generated by
fluctuations not accessible to mean-field theory.


Dispersionless branches of order-parameter susceptibilities which
lead to an infinite degeneracy of mean-field states, have been
found in a variety of models,\cite{FLUCTA,ROT,SLIDE,STACK} of
which perhaps the most celebrated is that in the
kagom\'e\cite{KAG} and pyrochlore\cite{PYRO} systems.  In almost
all cases, the dispersionless susceptibility is an artifact of
mean-field theory and does not represent a true symmetry of the
full Hamiltonian.  In such a case, the continuous degeneracy is
lifted by fluctuations, which can either be thermal
fluctuations\cite{VILL} or quantum fluctuations.\cite{EFS} Here we
have a rather unusual case in that the spin susceptibility has a
dispersionless direction (parallel to the inactive axis) which is
the result of an exact true symmetry of the quantum Hamiltonian
which persists even in the presence of thermal and quantum
fluctuations.

\section{LANDAU EXPANSION AT QUARTIC ORDER}

To discuss the nature of the ordered state one may consider the
self-consistent equations for the nonzero order-parameters which
appear below the ordering temperature at $kT_c=2\epsilon/3$ and
this is done in  III. However, the types of possible ordering
should also be apparent from the form of the anisotropy of the
free energy in order-parameter space which first occurs in terms
in the free energy which are quartic in the order-parameters.  In
principle, long-range order is only possible when we add to the
Hamiltonian terms which destroy the symmetry whereby one can
rotate arbitrarily planes of spins associated with a given orbital
flavor.  In the next section we study several perturbations which
stabilize long-range order.  Although the nature of the ordering
depends on the perturbation, generically the resulting dispersion
due to this symmetry-breaking perturbation  stabilizes the G
structure, so that the instabilities are confined to the
wavevector ${\bf q}={\bf Q}$. In this section we  implicitly
assume this scenario.

Accordingly, we now evaluate all terms in the free energy which
involve four powers of the critical variables
$\Bv^{\gamma}_{\alpha \alpha}({\bf Q})$ and
$\Av_{\alpha\alpha}({\bf Q})$ at the wavevector associated with
the assumed two sub-lattice, or G, structure.  These terms arise
from two mechanisms. The first contribution, which we denote
$F_4^{(4)}$, arises from ``bare'' quartic terms
in Eq. (\ref{XEQ}). The second type of contribution arises
indirectly through
$X^3(i)$
in Eq. (\ref{XEQ}).
There we have contributions to the free energy which involve
two critical variables and one noncritical variable (evaluated at
zero wavevector). When the free energy is minimized with respect
to this noncritical variable, we  obtain contributions to the
free energy which are quartic in the critical order-parameters and
which we denote $F_4^{(3)}$.

\subsection{Bare Quartic Terms, $F_4^{(4)}$}

The bare quartic terms are obtained from Eq. (\ref{bare4}), by
taking into account only diagonal matrix elements of the matrices
$\Av$ and $\vec{\Bv}$. Since the fourth-order term of the entropy
is multiplied by 18$kT$, [see Eq. (\ref{XEQ})], and we can safely
put here 18$kT$=12$\epsilon$, we find that the bare quartic terms
are given by
\begin{eqnarray}
F_4^{(4)}&=&24\epsilon\sum_{i}\sum_{\alpha}\Bigl [
\Av^{4}_{\alpha\alpha}(i)
+6\Av^{2}_{\alpha\alpha}(i)s_{\alpha}^{2}(i)+s_{\alpha}^{4}(i)\Bigr
],
\end{eqnarray}
where we have denoted
\begin{eqnarray}
s_{\alpha}^{2}(i)=(\Bv_{\alpha\alpha}^{x}(i) )^{2}+
(\Bv_{\alpha\alpha}^{y}(i) )^{2}+ (\Bv_{\alpha\alpha}^{z}(i)
)^{2}.\label{spin}
\end{eqnarray}
Introducing Fourier transformed variables via Eq. (\ref{FOURIER})
we thereby obtain terms quartic in the critical order parameters
as
\begin{eqnarray}
F_4^{(4)}& =&\frac{24\epsilon}{N}\sum_{\alpha}\Bigl
[\Av^{4}_{\alpha\alpha}
+6\Av_{\alpha\alpha}^{2}s_{\alpha}^{2}+s_{\alpha}^{4}\Bigr ],
\end{eqnarray}
where now all order parameters are to be evaluated at wavevector
${\bf Q}$. Using for the matrix elements of $\Av$ the
parametrization Eq. (\ref{map}), we find
\begin{eqnarray}
&&  F_4^{(4)} = {\epsilon \over N} \Bigl\{12( a_{1}^{2}+a_{2}^{2}
)^{2}
+48\sqrt{3}a_{1}a_{2} (s_{x}^{2}-s_{y}^{2})\nonumber\\
&+&48(a_{1}^{2}+a_{2}^{2} ) (s_{x}^{2}+s_{y}^{2}+s_{z}^{2} )+24
(s_{x}^{4}+s_{y}^{4}+s_{z}^{4} )\nonumber\\
&&\ \ \ \ \ \ -24 (a_{1}^{2}-a_{2}^{2} )
(s_{x}^{2}+s_{y}^{2}-2s_{z}^{2} ) \Bigr \}.
\end{eqnarray}

\subsection{Induced Quartic Terms, $F_4^{(3)}$}

To obtain the terms of this type, we first take from  Eq.
(\ref{bare3}) all the terms having diagonal matrix elements.
Multiplying them by $-6kT=-4\epsilon$ [see Eq. (\ref{XEQ})], we
have
\begin{eqnarray}
V_{3}&=&-8\epsilon\sum_{i\alpha}\Bigl
[\Av^{3}_{\alpha\alpha}(i)+3\Av_{\alpha\alpha}(i)\vec{\Bv}_{\alpha\alpha}(i)\cdot
\vec{\Bv}_{\alpha\alpha}(i)\Bigr ].
\end{eqnarray}
Next we insert here the Fourier transforms.  The critical
variables we treat here are the Fourier components at wavevector
${\bf Q} \equiv (\pi, \pi , \pi )/a $. When the wavevector is
${\bf Q}$, it will be left implicit.  We indicate explicitly only
those variables taken at zero wavevector.  Then  $V_3$ is given by
\begin{eqnarray}
V_{3}&=&-\frac{24\epsilon}{\sqrt{N}}\sum_{\alpha}\Bigl
[\Av_{\alpha\alpha}(0) (\Av_{\alpha\alpha}^{2}+s_{\alpha}^{2}
)
\nonumber\\
&+&2\vec{\Bv}_{\alpha\alpha}(0)\cdot\vec{\Bv}_{\alpha\alpha}
\Av_{\alpha\alpha}\Bigr ],
\end{eqnarray}
where we have used Eq. (\ref{spin}).

We now eliminate the noncritical variables at zero wavevector by
minimizing the free energy with respect to them.  We note that all the
noncritical zero wavevector variables have the same susceptibility
\begin{eqnarray}
\chi(0) = (12 kT + 8 \epsilon)^{-1} = (16 \epsilon)^{-1}  ,
\end{eqnarray}
and therefore the function to minimize is
\begin{eqnarray}
\tilde{V}_{3}&=&
V_{3}
+8\epsilon\sum_{\alpha}\Bigl
[\Av_{\alpha\alpha}^{2}(0)+\vec{\Bv}_{\alpha\alpha}(0)\cdot
\vec{\Bv}_{\alpha\alpha}(0)\Bigr ].\label{v3tilde}
\end{eqnarray}
The minimization procedure, allowing for the constraint
$\sum_{\alpha} \Av_{\alpha\alpha}(0)=0$, yields
\begin{eqnarray}
\Bv_{\alpha\alpha}^{\gamma}(0)&=&\frac{3}{\sqrt{N}}\Bv_{\alpha\alpha}^{\gamma}
\Av_{\alpha\alpha},\nonumber\\
\Av_{xx}(0)&=&\frac{1}{2\sqrt{N}}\Bigl
(2\Av_{xx}^{2}+2s_{x}^{2}-\Av_{yy}^{2}-s_{y}^{2}-\Av_{zz}^{2}-s_{z}^{2}\Bigr
),\nonumber\\
\Av_{yy}(0)&=&\frac{1}{2\sqrt{N}}\Bigl
(2\Av_{yy}^{2}+2s_{y}^{2}-\Av_{xx}^{2}-s_{x}^{2}-\Av_{zz}^{2}-s_{z}^{2}\Bigr
),\nonumber\\
\Av_{zz}(0)&=&-\Av_{xx}(0)-\Av_{yy}(0).
\end{eqnarray}
Inserting these values into Eq. (\ref{v3tilde}) yields  the
contribution $F_4^{(3)}$   to the  free energy
\begin{eqnarray}
&&F_4^{(3)}
=-\frac{72\epsilon}{N}\sum_{\alpha}\Av_{\alpha\alpha}^{2}s_{\alpha}^{2}-\frac{12\epsilon}{N}\Bigl
[\sum_{\alpha}(\Av_{\alpha\alpha}^{2}+s_{\alpha}^{2}
)^{2}\nonumber\\
&-&(\Av_{xx}^{2}+s_{x}^{2})(\Av_{yy}^{2}+s_{y}^{2})
-(\Av_{yy}^{2}+s_{y}^{2})(\Av_{zz}^{2}+s_{z}^{2})\nonumber\\
&&\ \ \ \ \ \ \ \ \ -
(\Av_{zz}^{2}+s_{z}^{2})(\Av_{xx}^{2}+s_{x}^{2})\Bigr ],
\end{eqnarray}
which, upon inserting the parametrization (\ref{map}) becomes
\begin{eqnarray}
F_4^{(3)} &=& {\epsilon \over N} \Bigl\{ - 12 \Bigl( \sum_\alpha
s_\alpha^2 \Bigr) ^2 + 36 \sum_{\alpha < \beta} s_\alpha^2
s_\beta^2 -3 (a_1^2+a_2^2)^2\nonumber\\
&-& 36 \sqrt 3 a_1a_2(s_x^2-s_y^2)   - 24 (a_1^2 + a_2^2) (s_x^2 +
s_y^2 + s_z^2) \nonumber\\
&& \ \ \ \ \ \ \ \ +18 (a_1^2 - a_2^2) (s_x^2 + s_y^2 - 2 s_z^2)
\Bigr \} .
\end{eqnarray}

\subsection{Total Fourth-Order Anisotropy}

Adding $F_4^{(3)}$ and $F_4^{(4)}$,
we find
$F_4$ as
\begin{eqnarray}
F_4  &=& {\epsilon \over N} \Bigl\{ 12 \Bigl( \sum_\alpha
s_\alpha^2 \Bigr)^2 -12 \sum_{\alpha < \beta} s_\alpha^2 s_\beta^2
+9 (a_1^2+a_2^2)^2\nonumber\\
&+&24 (a_1^2+a_2^2) \sum_\alpha s_\alpha^2 +12 \sqrt 3 a_1 a_2 [s_x^2 - s_y^2]\nonumber\\
&-&6 (a_1^2 - a_2^2) [s_x^2 + s_y^2 - 2 s_z^2]\Bigr\} \ ,
\label{FFOUR}
\end{eqnarray}
where all variables are evaluated at wavevector ${\bf Q}$. As
mentioned above, the anisotropy of this form determines the nature
of the mean-field states of the ideal KK Hamiltonian. We will give
a complete analysis of the symmetry and consequences of this
fourth order anisotropy in paper III. Here we will use this form
to determine the nature of possible ordered states in the presence
of symmetry-breaking perturbations such as the spin-orbit
interaction.

\section{SYMMETRY-BREAKING PERTURBATIONS}

As we have just seen, the idealized KK model considered above has
sufficient symmetry that there is no wavevector selection
\cite{WAVE} within mean-field theory and the exact symmetry of
this model does not support long-range order at nonzero
temperature.  In this section we consider the effects of various
additional perturbations which are inevitably present, even when
there is no distortion from perfect cubic symmetry.  We consider
in turn the effects of a) spin-orbit coupling, b) further neighbor
hopping, and c) Hund's rule or Coulomb exchange coupling. Here we
do {\it not} assume that the long-range order only involves the
wavevector ${\bf Q}$ of the G structure. In other words our first
objective is to see how these various perturbations lead to (if
they do) wavevector selection and what types of ordering result.

\subsection{Spin-Orbit Interactions}

We first consider the effect of including spin-orbit interactions,
since these interactions destroy the peculiar invariance with
respect to rotating planes of spins of different orbital flavors
independently. Below we see that the addition of spin-orbit
coupling leads to a wavevector selection from the susceptibility,
which previously had a dispersionless axis in the absence of such
a perturbation. Indeed, a plausible guess is that the system will
select the wavevector ${\bf Q}$ to allow simultaneous condensation
of spins of  the all three orbitals.

We write  the spin-orbit interaction, $V_{\rm SO}$, as
\begin{eqnarray}
V_{\rm SO} &=& \lambda \sum_i \sum_{\alpha \beta \gamma} \sum_{\mu
\nu} \langle \alpha | L_\gamma | \beta \rangle c_{i \alpha
\mu}^\dagger c_{i \beta \nu} [ \sigmav^\gamma]_{\mu \nu} ,
\label{SPINORBIT}
\end{eqnarray}
where
\begin{eqnarray}
\langle \alpha | L_\gamma | \beta \rangle = -i \epsilon_{\alpha
\gamma \beta} ,
\end{eqnarray}
and  $\lambda$ is the spin-orbit coupling constant.
We now incorporate this perturbation into the mean-field
treatment. The expression for the entropy does not need to be
changed.  The trial energy involves ${\rm Tr} \left[ \rhov(i)
V_{\rm SO} \right]$ and generates a perturbative contribution to
the free energy which is
\begin{eqnarray}
\delta F &=& 2 \lambda \sum_i \sum_{\alpha \beta \gamma}
\Bv_{\alpha\beta}^{\gamma}(i)\langle \beta | L_\gamma | \alpha
\rangle .
\end{eqnarray}
In terms of Fourier transformed variables this is
\begin{eqnarray}
\delta F &=& 2 \lambda N^{1/2} \sum_{\alpha \beta \gamma}
\Bv_{\alpha \beta }^{\gamma} (q=0) \langle \beta | L_\gamma |
\alpha \rangle . \label{FREELAMBDA}
\end{eqnarray}
Thus the spin-orbit interaction
appears as a field acting on the noncritical order-parameter
$\vec{\Bv}_{\alpha\beta}(q=0)$, with $\alpha\neq\beta$.

We now calculate the perturbative effect of the spin-orbit
interaction. Because the perturbation $V_{\rm SO}$ is the only
term in the Hamiltonian that causes a transition from one orbital
to another, the leading perturbation to the free energy will be of
order $\lambda^2$. We develop an expansion at temperatures
infinitesimally below $T_c=2\epsilon/(3k)$ in powers of $\lambda$
and $\{\psi \}$, where $\{\psi\}$ denotes the set of variables
which, in the absence of spin-orbit coupling, are critical at the
highest temperature, namely, $kT=2 \epsilon/3$. This set includes
$\Bv_{\alpha \alpha}^{\gamma}({\bf q})$ for ${\bf q}$ on its
``soft line'',  which is $q_\alpha$ arbitrary and the other
components equal to $\pi/a$. In addition, this set also includes
$\Av_{\alpha\alpha}({\bf Q})$, namely, $a_1({\bf Q})$ and
$a_2({\bf Q})$. The dominant perturbation to the free energy will
be of order $\lambda^2 \psi_i\psi_j$, where $\psi_i$ is one of the
critical order parameters. Terms of order $\lambda^2 \psi_i$ are
not allowed, as they would cause ordering at all temperatures
above $T_c$ and contributions independent of $\psi_i$ are of no
interest to us. So our goal is to calculate all terms of order
$\lambda^2 \psi_i\psi_j$.  By modifying the terms quadratic in the
critical order parameters we will obtain a free energy without a
dispersionless branch of the susceptibility, and therefore the
spin-orbit perturbation will lead to wavevector selection.

Terms of order $\lambda^2 \psi_i\psi_j$ in the free energy arise
from either bare fourth-order terms or indirectly from cubic terms
which involve one noncritical variable and two critical variables.
Here we describe these contributions qualitatively.  The explicit
calculations are given in Appendices \ref{B} and \ref{C}.  We
first consider contributions arising from the third-order terms.
Note that the spin-orbit perturbation $V_{\rm SO}$ acts like a
``field'' in that it couples linearly to the order parameter
$\Bv_{\alpha \beta }^{\gamma} ( {\bf q}=0)$, as one can see from
Eq. (\ref{FREELAMBDA}). Minimization with respect to this order
parameter yields
\begin{eqnarray}
\Bv_{\alpha \beta}^{ \gamma} ( {\bf q}=0) &=& - {\lambda \over 6
\epsilon} N^{1/2} \langle \beta | L_\gamma | \alpha \rangle
\equiv  i N^{1/2} g_0 \epsilon_{\alpha \beta \gamma}  ,
\label{GEQ}
\end{eqnarray}
where $g_0= \lambda/(6 \epsilon)$ and we noted   that the
non-diagonal inverse susceptibility $\chi^{-1}_{\alpha \beta}(0) $
is $12 \epsilon$ at $kT=2\epsilon/3$ [see Eq. (\ref{chind})]. In
other words, we have the spatially uniform displacement,
$\Bv_{\alpha \beta}^{ \gamma} (i) = i g_0 \epsilon_{\alpha \beta
\gamma}$, which is linear in $\lambda$. Now consider third-order
terms in the free energy which are schematically of the form
\begin{eqnarray}
\delta F &=& a \Bv_{\alpha \beta}^{ \gamma}({\bf q}=0) \psi_i x_j
, \label{CUBIC}
\end{eqnarray}
where $a$ is a constant, and $x_j$ is a noncritical variable, so
that its susceptibility $\chi_j$ is finite at $T_c$. After
minimizing with respect to $x_j$, we  obtain a contribution to
the free energy of order $-(1/2) \chi_j a^2 (\Bv_{\alpha \beta}^{
\gamma}({\bf q}=0))^2 \psi_i^2$, which is a term of order
$\lambda^2 \psi_i \psi_j$ (albeit with $i=j$). This perturbative
contribution to the free energy quadratic in the critical
variables will be denoted $F_2^{(3)}$. Note that these cubic terms
[see Eq. (\ref{CUBIC})] are identified as being linear in (a)
$\Bv_{\alpha \beta}^{ \gamma}$,
in (b) a critical order-parameter $\psi_i$, such as
$\vec{\Bv}_{\alpha\alpha} ( q_\alpha)$ (by this we mean
$\vec{\Bv}_{\alpha\alpha}$ evaluated for a wavevector on its soft
line), or $\Av_{\alpha\alpha}({\bf Q})$, and in (c) some
noncritical order-parameter. Terms of order $\lambda^2 \psi_i
\psi_j$ can also come from bare fourth order terms which are
products of two powers of $\Bv_{\alpha \beta}^{ \gamma}({\bf
q}=0)$ with two critical variables and these contributions are
denoted $F_2^{(4)}$.  All these terms will then lead to
modifications of the terms in the free energy which are quadratic
in the critical variables and which therefore may lead to
wavevector selection within the previously dispersionless critical
sector.

We now identify cubic terms in Eq. (\ref{XEQ}) which are of the
form written in Eq. (\ref{CUBIC}).  There are no nonzero cubic
terms which are linear in both $\lambda$ and either $a_1({\bf Q})$
or $a_2({\bf Q})$.  The allowed cubic terms
are analyzed in
Appendix \ref{B} and the result for their perturbative
contribution $\delta F_2^{(3)}$ to the free energy from minimizing
these cubic terms is
\begin{eqnarray}
F_2^{(3)} &=& - C_0 \sum_{\alpha \beta \gamma}\epsilon_{\alpha
\beta \gamma}^2 \Bigl\{ \sum_{q_\alpha} \Bigl( 2 s_{\alpha
\gamma}({\bf q}) s_{\alpha \gamma}(-{\bf q})\nonumber\\
& +& s_{\alpha \alpha}({\bf q}) s_{\alpha \alpha}(-{\bf q})
\Bigr)\nonumber\\
 &+& \Bigl( s_{\alpha \gamma}({\bf Q})s_{\beta \gamma}({\bf Q}) - 2
s_{\beta \beta} ({\bf Q}) s_{\alpha \beta}({\bf Q}) \Bigr) \Bigr\}
,\label{ff3}
\end{eqnarray}
where $C_0=144 g_0^2 \epsilon =4 \lambda^2/\epsilon$, and we have
introduced the definition
\begin{eqnarray}
s_{\alpha\beta}({\bf q})=\Bv_{\alpha\alpha}^{\beta}({\bf
q}).\label{spindef}
\end{eqnarray}
In Eq. (\ref{ff3}),  $\sum_{q_\alpha}$ means that the wavevector
is summed over the soft line so that $q_\mu=\pi /a$ for $\mu \not=
\alpha$ and $q_\alpha$ ranges from $-\pi /a$ to $\pi/a$.  In
particular the sum over $q_\alpha$ also includes ${\bf q}={\bf
Q}$.  In Appendix \ref{C} we evaluate the bare quartic terms in
the free energy which also give a result of order $\lambda^2
\psi_i \psi_j$, and find
\begin{eqnarray}
&& F_2^{(4)} = C_0 \Bigl\{ {4 \over 3} \sum_{\alpha \gamma}
\sum_{q_\alpha} s_{\alpha \gamma}({\bf q}) s_{\alpha \gamma}(-{\bf
q}) +a_1^{2}({\bf Q}) + a_2^{2}({\bf Q})\nonumber \\ &&  + {1
\over 3} \sum_{\alpha \beta \gamma} \epsilon_{\alpha \beta
\gamma}^2 \Bigl( 2 s_{\alpha \gamma}({\bf Q}) s_{\beta
\gamma}({\bf Q}) - \sum_\nu s_{\alpha \nu}({\bf Q}) s_{\beta \nu}
({\bf Q}) \Bigr) \Bigr\}
.\nonumber\\
\end{eqnarray}

We now discuss the meaning of these results.  One
effect of the spin-orbit contributions is to couple critical spin variables
of different orbitals.  But this type of coupling only takes
place at the wavevector ${\bf Q}$ at which spin variables for both
orbitals are simultaneously critical.  So we write the sum of all
the quadratic perturbations in terms of spin variables
$s_{\alpha \gamma}$ listed above as
\begin{eqnarray}
\delta F_2 &=& \frac{1}{2}\sum_\alpha \sum_{\mu \nu} \Bigl[ \Bigl(
\sum_{q_\alpha} [M_{\rm d}^{(\alpha)} ]_{\mu  \nu} s_{\mu
\alpha}({\bf q}) s_{\nu \alpha}(-{\bf q}) \Bigr) \nonumber \\ && \
+ [M_{\rm o}^{(\alpha)} ]_{\mu \nu} s_{\mu \alpha}({\bf Q}) s_{\nu
\alpha}({\bf Q}) \Bigr]  ,
\end{eqnarray}
where ${\bf M}_{\rm d}^{(\alpha)}$ is a diagonal matrix and
${\bf M}_{\rm o}^{(\alpha)}$ is an off-diagonal matrix.
These matrices are
\begin{eqnarray}
{\bf M}_{\rm d}^{(\alpha)} &=& - {4C_0 \over 3} \left[
\begin{array} {c c c }
1 &  0 &  0 \\
0 &  1 &  0 \\
0 &  0 &  1 \\ \end{array} \right],
\nonumber \\
{\bf M}_{\rm o}^{(\alpha)} &=& {4 C_0 \over 3} \left[
\begin{array} {c c c }
 0 &  1 &  1 \\
 1 &  0 & -1 \\
 1 & -1 &  0 \\ \end{array} \right]  ,
\label{EQ63}
\end{eqnarray}
where the first row and column refers to $s_{\alpha\alpha}$ and
the other two refer to $s_{\beta \alpha}$, with $\beta \not=
\alpha$. The contributions to the free energy from ${\bf M}_{\rm
d}^{(\alpha)}$ are independent of wavevector and thus do not
influence wavevector selection.  The term in ${\bf M}_{\rm
o}^{\alpha}$ selects ${\bf Q}$ (because the minimum eigenvalue of
the matrix ${\bf M}_{\rm o}^{(\alpha)}$ is $-4C_0/3$, which is
negative). In addition, the minimum eigenvector determines the
linear combination of order parameters that is critical.  If this
eigenvector has components $(c_1, c_2, c_2)$, then, for
$\alpha=x$, we have
\begin{eqnarray}
s_{xx}({\bf Q}) =  \xi_x c_1 \ ,\
s_{yx}({\bf Q}) =  \xi_x c_2 \ ,\
s_{zx}({\bf Q}) = \xi_x  c_2  ,
\end{eqnarray}
where $\xi_x$ is the normal mode amplitude and we adopt the
normalization $c_1^2 + 2 c_2^2=1$.  Thus, out of the nine spin
components $s_{\alpha \beta}({\bf Q})$ which were simultaneously
critical in the absence of spin-orbit coupling, we have the spin
fluctuation corresponding to the three normal-mode amplitudes
$\xi_x$, $\xi_y$, and $\xi_z$ in terms of which we write the
staggered spin vector for orbital $\alpha$, ${\bf s}_{\alpha}({\bf
Q})$, as
\begin{eqnarray}
{\bf s}_x({\bf Q})  &\equiv & (s_{xx}({\bf Q}),s_{xy}({\bf
Q}),s_{xz}({\bf Q}))
= (\xi_x c_1, \xi_y c_2 , \xi_z c_2), \nonumber \\
{\bf s}_y({\bf Q}) &=& (\xi_x c_2, \xi_y c_1 , \xi_z c_2), \nonumber \\
{\bf s}_z({\bf Q}) &=& (\xi_x c_2, \xi_y c_2 , \xi_z c_1)  .
\end{eqnarray}
The {\it total} spin at site $i$ is the sum of the spins associated
with each orbital flavor and is given by the staggered spin vector
\begin{eqnarray}
{\bf S}({\bf Q})  &=& (\xi_x, \xi_y, \xi_z)(c_1+2c_2) \ ,
\end{eqnarray}
so that the $\xi$'s are proportional to the components of the
total spin. Now we evaluate the fourth-order free energy terms
relevant to the spin order-parameters [see Eq. (\ref{FFOUR})] in
terms of these critical order parameters $\xi_i$:
\begin{eqnarray}
\delta F &=& C_1 \Bigl( [\xi_x^2 + \xi_y^2 +
\xi_z^2]^2[c_{1}^{4}+3c_2^4+2c_1^2c_2^2] \nonumber \\ && \ -
[\xi_y^2\xi_z^2 + \xi_x^2\xi_z^2 + \xi_x^2\xi_y^2]
[c_1^2-c_2^2]^{2} \Bigr) \ , \label{QUARTIC}
\end{eqnarray}
where $C_1$ is a constant.  In general, a form like this would
have ``cubic'' anisotropy in that the vector $\xiv$ (the total
spin vector) would preferentially lie along a $(1,1,1)$ direction
in order to maximize the negative term in $\xi_\alpha^2
\xi_\beta^2$. However, for the present case, the minimum
eigenvector of ${\bf M}_{\rm o}^{(\alpha)}$ is $(c_1, c_2, c_2)
\propto (1,-1,-1)$. Thus for the present case $c_1^2=c_2^2$, and
the quartic term is isotropic in $\xi$ space. What this means is
that although the spin-orbit interaction selects the directions
for the spin vectors ${\bf s}_{\alpha}$ of orbital flavor $\alpha$
relative to one another, there is rotational invariance when all
the ${\bf s}_\alpha$'s are rotated together. This indicates that
relative to the mean-field state there are zero frequency
excitations which correspond to rotations of the staggered spin.
Here we find this result at order $\lambda^2$.  More generally,
one can establish this rotational invariance to all orders in
$\lambda$ and without assuming the validity of mean-field
theory.\cite{PRL,TY1}

Note that the spin state induced by spin-orbit coupling (with
$c_1=-c_2$) does {\it not} have the spins of the individual orbitals,
${\bf s}_\alpha$, parallel to one another and thus the net spin,
${\bf S}$, is greatly reduced by this effect. Explicitly, when
$c_1=-c_2$, we have
\begin{eqnarray}
S^2 &=& (\xi_x^2+\xi_y^2+\xi_z^2) c_1^2 \nonumber\\
&=& {\bf s}_{x}^{2}({\bf Q})
= {\bf s}_{y}^{2}({\bf Q}) = {\bf s}_{z}^{2}({\bf Q}) = (\xi_x^2+
\xi_y^2+\xi_z^2)/3 .
\end{eqnarray}
This means that the total spin squared is 1/3 of what it would be if
the ${\bf s}_\alpha$ were parallel to one another.

It remains to check that the variables $a_k({\bf Q})$ are less
critical than $s_{\alpha \gamma}({\bf Q})$.  The results given in
Eq. (\ref{QUADORB}) of Appendix \ref{C} show a positive shift in
the free energy associated with the variables $a_k({\bf Q})$,
whereas the spin variables have a negative shift in free energy
due to spin-orbit interactions. We therefore conclude that in the
presence of spin-orbit interactions, mean-field theory does give
wavevector selection and one has the usual two-sublattice
antiferromagnet, but with a greatly reduced spin magnitude.  It is
interesting to note that\cite{LTO} LaTiO$_3$ has a zero point
moment which is about 45\% of the value of the spin were fully
aligned.  This zero-point spin reduction is much larger than would
be expected for a conventional spin 1/2 Heisenberg system in three
spatial dimensions.  It is possible that spin-orbit interactions
might partially explain this anomalous spin reduction.

\subsection{Further Neighbor Hopping}

\begin{figure}
\leavevmode \epsfclipon \epsfxsize=6.truecm
\hspace{1cm}\vbox{{\epsfbox{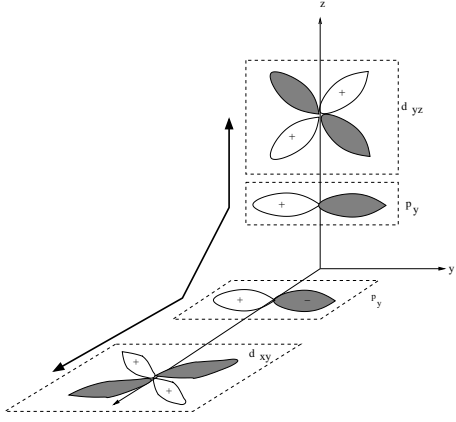}}
}
\vspace{.5cm} \caption{Hopping between different orbitals on
next nearest neighboring (nnn) Ti ions when hopping between
neighboring oxygen p orbitals is allowed. The hopping matrix element
is the product of matrix elements to hop from a Ti ion in a
$d_{yz}$ state to an O ion in a $p_y$ state, then to an adjacent O
ion also in a $p_y$ state, and finally to a nnn Ti ion in a
$d_{xy}$ state.  
} 
\label{NNNHOP}
\end{figure}

\begin{figure}
\leavevmode \epsfclipon \epsfxsize=6.truecm
\hspace{1cm}\vbox{\epsfbox{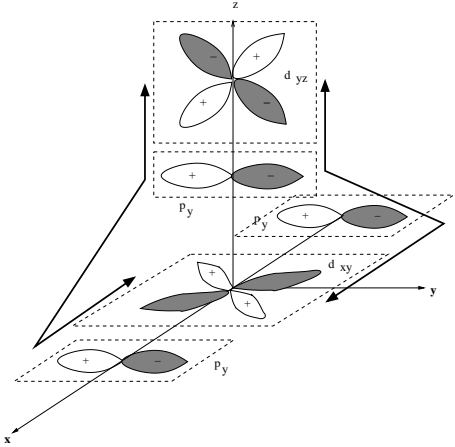}} \vspace{.5cm}
\caption{Hopping between different orbitals on nearest-neighboring
Ti ions when hopping between neighboring oxygen p orbitals is
allowed. The matrix elements for the two channels to hop from
$d_{yz}$ to $d_{xy}$ have opposite signs, so that the total matrix
element (summed over the two channels) is zero, as one would
deduce from symmetry considerations. Thus the only processes
involving two nearest neighboring oxygen ions are processes like
those shown in Fig. 3 between nnn Ti ions.} \label{NNNHOP2}
\end{figure}

We now consider the effect of adding nnn hopping to the Hubbard model
of Eq. (\ref{HHUB}).  For a perfectly cubic system, this
hopping process comes from the next-to-shortest exchange path between
magnetic ions, as is shown in Fig. \ref{NNNHOP}.  We write the
perturbation $V$ to the Hubbard Hamiltonian due to these processes as
\begin{eqnarray}
V = t' \sum_\alpha \gamma_\alpha(i,j) V_{ij} ,
\end{eqnarray}
where $t'$ is the effective hopping matrix element connecting next-nearest
neighbors, $\alpha$ is summed over coordinate directions $x$, $y$, and $z$,
$\gamma_\alpha(i,j)$ is unity if sites $i$ and $j$ are next-nearest
neighbors in the same $\alpha$-plane and is zero otherwise, and
\begin{eqnarray}
V_{ij} = \sum_\sigma \sum_{\beta \delta} \epsilon_{\alpha \beta
\delta}^2 c_{i \beta \sigma}^\dagger c_{j \delta \sigma} .
\end{eqnarray}
Here $\alpha$ is in the direction normal to the plane containing
spins $i$ and $j$, and $\epsilon_{\alpha \beta \delta}^2$
restricts the sum over $\beta$ and $\delta$ to the two ways of
assigning indices so that $\alpha$, $\beta$, and $\delta$ are all
different. 
Note that the paths from $i\beta$ to $j\delta $ and from
$i\delta$ to $j\beta$ use alternate paths of the square plaquette
connecting $i$ and $j$.
Notice that the processes which couple nearest
neighbors cancel by symmetry (see Fig. \ref{NNNHOP2}), so that the
effect of hopping between magnetic ions via two intervening oxygen
ions involves only nnn hopping.  This
generates a perturbation to the KK
Hamiltonian (which describes the low-energy manifold) of the form
\begin{eqnarray}
V_{KK} &=& - \epsilon' \sum_\alpha \sum_{ij}
\gamma_\alpha(i,j)\nonumber\\
&&\times \Bigl( \sum_{\beta \delta \sigma} \epsilon_{\alpha \beta
\delta}^2 c_{i\beta\sigma}^\dagger c_{j\delta\sigma} \Bigr) \Bigl(
\sum_{\beta \delta \sigma} \epsilon_{\alpha \beta \delta}^2
c_{j\beta\sigma}^\dagger c_{i\delta\sigma} \Bigr) ,\label{VKK}
\end{eqnarray}
where $\epsilon'=(t')^2/U$ and $U$ is the on-site Coulomb energy.
This may be written as
\begin{eqnarray}
V_{KK} &=& \epsilon' \sum_\alpha \sum_{ij} \gamma_\alpha(i,j)
V_\alpha(i,j)  ,
\end{eqnarray}
where, apart from a term which is a constant in the low-energy manifold,
we have for $\alpha=x$
\begin{eqnarray}
V_x(i,j) &=&
\sum_{\sigma \eta}\Bigl(
c_{iy\sigma}^\dagger c_{iz\eta} c_{jy\eta}^\dagger c_{jz\sigma} +
c_{iy\sigma}^\dagger c_{iy\eta} c_{jz\eta}^\dagger c_{jz\sigma}
\nonumber \\ && \ + c_{iz\sigma}^\dagger c_{iz\eta}
c_{jy\eta}^\dagger c_{jy\sigma} + c_{iz\sigma}^\dagger c_{iy\eta}
c_{jz\eta}^\dagger c_{jy\sigma} \Bigr)  ,
\end{eqnarray}
and similarly for $y$ and $z$.

The details of the mean-field treatment of this perturbation is
given in Appendix \ref{D}. Here we summarize the major analytic
results obtained there for the wavevector-dependent spin
susceptibility at the critical wavevector, ${\bf Q}$,
$\chi_{\alpha \sigma ; \beta \sigma'} ({\bf Q})= \chi_{\alpha
\beta } ({\bf Q}) \delta_{\sigma , \sigma'}$, where $\alpha$ and
$\beta$ are orbital indices and $\sigma$ and $\sigma'$ are spin
indices.  The result of Appendix \ref{D} is that
\begin{eqnarray}
\chi_{\alpha \beta} ({\bf Q})^{-1} &=&
\left[ \begin{array} {c c c}
12 kT - 8 \epsilon  & 8 \epsilon^\prime & 8 \epsilon^\prime \\
8 \epsilon^\prime & 12kT - 8 \epsilon & 8 \epsilon^\prime  \\
8 \epsilon^\prime & 8 \epsilon^\prime & 12kT - 8 \epsilon \\
\end{array} \right]  .
\end{eqnarray}
The minimum eigenvalue is
\begin{eqnarray}
\lambda &=& 12 kT - 8 \epsilon - 8 \epsilon^\prime  .
\end{eqnarray}
This gives
\begin{eqnarray}
kT_c &=&  2 (\epsilon + \epsilon^\prime)/3  .
\end{eqnarray}
By considering the eigenvectors and the effect of the fourth order
terms, the analysis of Appendix \ref{D} shows that nnn hopping
does stabilize a ${\bf Q}$ antiferromagnetic structure, but the
resulting 120$^{\rm o}$ state has zero net staggered spin.  In
addition, as before, there is a degeneracy between the spin-only
states we have just described, and a state involving orbital
order. As shown in III, fluctuations remove this degeneracy, so
that we may consider only the mean-field solutions for spin-only
states.  Such a magnetic structure for which the local moment
(summed over all flavors) vanishes, will be rather difficult to
detect experimentally.

It is instructive to argue for the above results without actually
performing the detailed  calculations of Appendix \ref{D}.  We
expect the effect of indirect exchange between nnn's to induce an
antiferromagnetic interaction between the spins of {\it different}
orbital flavors of nnn's.  Note that the wavevector ${\bf Q}$
describes a two sub-lattice structure in which  nnn's are on the
{\it same} sub-lattice.  Accordingly, as far as mean-field theory
is concerned, an nnn interaction between different flavors is
equivalent to an antiferromagnetic interaction between spins of
different flavors on the same site.  So the spins of the three
orbital flavors form the same structure as a triangular lattice
antiferromagnet,\cite{TAF} namely the spins of the three different
orbital flavors are  equal in magnitude and  all lie in a
single plane with orientations 120$^{\rm o}$ apart.  This state
still has  global rotational invariance, but also, as does
the triangular lattice antiferromagnet, it has  degeneracy
with respect to rotation of the spins of two flavors about the axis
of the spin of the third flavor.

\subsection{Hund's Rule Coupling}

We now consider the effect of Hund's rule coupling.  Our aim is to
see how this perturbation selects an ordered phase from among
those phases which would first become critical in the absence of
this perturbation as the temperature is reduced. To leading order
in $\eta \equiv J_H/U$, where $J_H$ is the Hund's rule coupling
constant (which is positive in real systems), as discussed in
Appendix \ref{E}, this perturbation reads \cite{IHM}
\begin{eqnarray}
\delta {\cal H}_{KK}&=&\epsilon\eta\sum_{\langle
ij\rangle}\sum_{\beta\gamma\neq\langle
ij\rangle}\sum_{\sigma\sigma '}\Bigl
(c^{\dagger}_{i\gamma\sigma}c_{i\beta\sigma}c^{\dagger}_{j\gamma\sigma
'}c_{j\beta\sigma '}\nonumber\\
&-&c^{\dagger}_{i\gamma\sigma '
}c_{i\beta\sigma}c^{\dagger}_{j\gamma\sigma }c_{j\beta\sigma '}+
c^{\dagger}_{i\gamma\sigma}c_{i\beta\sigma}c^{\dagger}_{j\beta\sigma
'}c_{j\gamma\sigma '}\nonumber\\
&-& c^{\dagger}_{i\beta\sigma '
}c_{i\beta\sigma}c^{\dagger}_{j\gamma\sigma }c_{j\gamma\sigma '}
-2c^{\dagger}_{i\beta\sigma}c_{i\beta\sigma}c^{\dagger}_{j\gamma\sigma
'}c_{j\gamma\sigma '}\nonumber\\
&+&2c^{\dagger}_{i\gamma\sigma '
}c_{i\beta\sigma}c^{\dagger}_{j\beta\sigma }c_{j\gamma\sigma
'}\Bigr ),
\end{eqnarray}
where $\epsilon=t^2/U$, as before. \cite{rem} To see the effect of
this perturbation within mean-field theory, we calculate its
average (see Appendix \ref{E} for details). Confining  to averages
which are critical when $\eta=0$, (i.e., $\Av_{\alpha\alpha}$ and
$\vec{\Bv}_{\alpha\alpha}$), the result of Appendix \ref{E} is
\begin{eqnarray}
\langle \delta {\cal H}_{KK}\rangle &=&\epsilon\eta\sum_{\langle
ij\rangle}\sum_{\beta\gamma\neq\langle ij\rangle}\Bigl
(10\Av_{\beta\beta}(i)\Av_{\beta\beta}(j)\nonumber\\
&-&10\Av_{\beta\beta}(i)\Av_{\gamma\gamma}(j)
+2\vec{\Bv}_{\beta\beta}(i)\cdot\vec{\Bv}_{\beta\beta}(j)\nonumber\\
&-&
2\vec{\Bv}_{\beta\beta}(i)\cdot\vec{\Bv}_{\gamma\gamma}(j)\Bigr ).
\end{eqnarray}
Using Eqs. (\ref{map}) and (\ref{spindef}) to write the order
parameters in terms of the $a_{\ell}$'s and the
$s_{\alpha\gamma}$'s, this contributes a perturbation to the free
energy given by
\begin{eqnarray}
\delta F &=& \frac 12 \sum_{k,l} \delta \left[ \chiv_{n}^{-1}
({\bf
q}) \right]_{kl}  a_k({\bf q}) a_l ( - {\bf q})\nonumber\\
& +&\frac{1}{2} \sum_{\alpha \beta \gamma} \delta \left[
\chiv_{s}^{-1} ({\bf q})\right]_{\alpha \beta}^\gamma s_{\alpha
\gamma}({\bf q}) s_{\beta \gamma}(-{\bf q})   , \label{EQ88}
\end{eqnarray}
where
\begin{eqnarray}
&&\delta [ \chiv_{n}^{-1} ( {\bf q}) ] \nonumber\\
  &=&
-20\epsilon\eta\left
[\begin{array}{cc}-\frac{1}{3}(2c_{x}+2c_{y}-c_{z})&\frac{1}{\sqrt{3}}(c_{x}-c_{y})\\
\frac{1}{\sqrt{3}}(c_{x}-c_{y})&-c_{z}\end{array}\right],
\end{eqnarray}
and
\begin{eqnarray}
\delta [ \chiv _{s}^{-1}( {\bf q}) ] &=&-4\epsilon\eta\left
[\begin{array}{ccc}0&c_{z}&c_{y}\\
c_{z}&0&c_{x}\\
c_{y}&c_{x}&0\end{array}\right ].\label{EQ90}
\end{eqnarray}
If the minimum eigenvalue of $\delta \chiv^{-1}$ at wavevector
${\bf Q}$ is negative, then the instability temperature for the
associated order parameter is raised by the perturbation and vice
versa. Note that at wavevector ${\bf Q}$, $c_x=c_y=c_z=-1$ the
eigenvalues of $\delta \left[ \chiv_s^{-1} ({\bf q}) \right]$ are
$8 \eta \epsilon$, $-4\eta \epsilon$, and $-4\eta \epsilon$. On
the other hand, the eigenvalues of $\delta \left[ \chiv
^{-1}_n({\bf q}) \right]$ are  both $-20 \eta \epsilon$.  From
this result we conclude that Hund's rule coupling favors
antiferromagnetic orbital ordering, as described by the order
parameters $a_1({\bf Q})$ and $a_2({\bf Q})$.  Since the
mean-field temperature for spin and orbital ordering were
degenerate for $\eta=0$, we conclude that within mean-field theory
the addition of an infinitesimal Hund's rule coupling gives rise
to an ordering transition in which the ordered state shows
long-range antiferromagnetic orbital order, characterized by the
order-parameters $a_1({\bf Q})$ and $a_2({\bf Q})$.  However,
since we have shown elsewhere\cite{IV} that for the bare KK model,
fluctuations stabilize the spin-only states relative to orbital
states, we conclude that when fluctuations are taken into account,
it will take a finite amount of Hund's rule coupling to bring
about orbital ordering. For spin ordering the mean-field state is
degenerate with respect to an arbitrary rotation.  This is
reflected by the fact that the term which is fourth order in the
spin components is isotropic.

We now discuss the anisotropy in the mean-field solution for
orbital order.  We want to determine the form the free energy
assumes in terms of the Fourier-transformed variables $a_1({\bf
Q})$ and $a_2({\bf Q})$.  Wavevector conservation dictates that we
can have only products involving an even number of these
variables. If we write $a_1({\bf Q})=a\cos \theta_{\bf Q}$ and
$a_2({\bf Q})=a \sin \theta_{\bf Q}$, then we show in Appendix
\ref{F} that the contribution to the free energy of order $a^4$ is
independent of $\theta_{\bf Q}$, but the term of order $a^6$ is of
the form $\delta F = a^6 [ C_0 + C_6 \cos (6 \theta_{\bf Q} +
\phi)]$.
This form indicates an anisotropy, so that the mean-field solution
is not subject to a rotational degeneracy in $a_1$-$a_2$ space. If
$C_{6}$ is positive and $\phi =0$, these minima come from the six
angles that are equivalent to $\theta_{\bf Q}=\pi/2 + n\pi/3$. For
$\theta_{\bf Q}=\pi/2$, $a_1=0$ and we have ordering involving
only $a_2$, so that $\langle N_z \rangle=1/3$, $\langle N_x
\rangle = 1/3 + \sqrt 2 a_2(i)$ and $\langle N_y \rangle = 1/3 -
\sqrt 2 a_2(i)$. The six minima of $\cos (6 \theta_{\bf Q})$
correspond to the six permutations of coordinate labels which give
equivalent ordering under cubic symmetry. Somewhat different
states occur for $C_{6}$ negative, but different solutions
reproduce the cubic symmetry operations.

\subsection{Spin-Orbit Interactions and Hund's Rule Coupling}

Here we briefly consider the case when we include the effects of
both spin-orbit and Hund's rule coupling.  We consider the
instabilities at wavevector ${\bf Q}$.  In this case we construct
the spin susceptibility $\chiv^{-1}_s({\bf Q})$ [defined as in Eq.
(\ref{EQ88})]. For the present case we may use our previous
calculations in Eqs. (\ref{EQ63}) and (\ref{EQ88}) to write
\begin{eqnarray}
\chiv_s^{-1}(\gamma) &=&  \left[ \begin{array} {c c c}
\lambda_0 + x & y & y \\
y & \lambda_0 + x & z \\
y & z & \lambda_0 + x \\
\end{array} \right]  ,
\end{eqnarray}
where the first row and column refer to $s_{\gamma \gamma}$ and the other
two rows and columns refer to $s_{\beta \gamma}$ with $\beta \not= \gamma$
and
\begin{eqnarray}
x = - \frac {4}{3} C_0  , \  \ y = \frac 43 C_0 + 4 \epsilon\eta
,\ \ z=-\frac{4}{3}C_{0}+4\epsilon\eta .
\end{eqnarray}
Similarly the orbital susceptibility (also at wavevector ${\bf Q}$) is
given by
\begin{eqnarray}
\chiv({\bf Q})^{-1}_n &=&  \left[ \begin{array} {c c}
\lambda_0 + w & 0 \\
0 & \lambda_0 + w \\
\end{array} \right]  ,
\end{eqnarray}
where
\begin{eqnarray}
w &=& 2C_0 - 20  \epsilon \eta \ .
\end{eqnarray}
In the above $C_0= 4 \lambda^2/\epsilon$ must be positive,
$\lambda_{0}=12kT+8\epsilon$,  and $\eta\equiv J_H/U$ is normally
positive, although we may draw a phase diagram incorporating the
possibility that $\eta$ is negative.

As we have seen, with only spin-orbit interactions we get a spin state
which has a rotational degeneracy, and with only Hund's rule interactions,
the ordered phase has orbital rather than spin ordering.  When both interactions
are present, there is a competition between these two types of ordering.
To study this competition we need to compare the minimum  eigenvalue of
the two susceptibility matrices given above.  For the inverse spin
susceptibility matrix $y \geq z$, in which case the minimum eigenvalue is
\begin{eqnarray}
\lambda_- &=& \lambda_0 + x + (z/2) - \sqrt{ (z/2)^2 + 2 y^2} \ .
\end{eqnarray}
On dimensional grounds, we expect that for $C_0 < \tau \eta
\epsilon$, where $\tau$ is a constant,  Hund's rule coupling will
dominate and will lead to orbital ordering. Indeed after some
algebra we find this condition with $\tau\approx 2.7$. This  may
be written as $\eta>0$ and $\lambda < \tau'\epsilon \sqrt {\eta
}$, where $\tau'= \sqrt \tau /2 \approx 0.82$

\begin{figure}
\leavevmode \epsfclipon \epsfxsize=6.truecm
\hspace{1cm}\vbox{\epsfbox{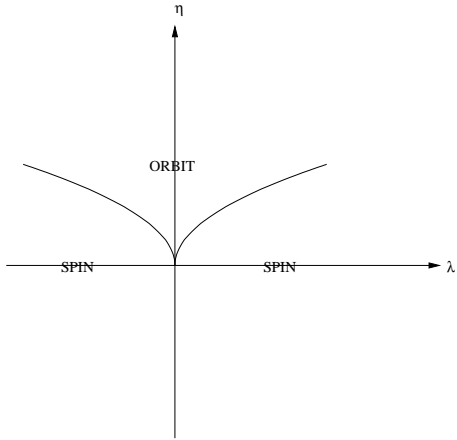}} \vspace{.5cm} 
\caption{The
mean-field phase diagram as a function of the spin-orbit coupling
constant $\lambda$ and the Hund's rule coupling constant
$\eta\equiv J_H/U$ (which is normally positive).  In the
``spin-only'' phase for $\eta \not= 0$, the staggered moment
orients along a $(1,1,1)$ direction, but the staggered spin
moments of different orbital states are not collinear, thus
reducing the net staggered spin. For $\eta=0$, the mean-field
state has rotational degeneracy, so no easy direction of staggered
magnetization is selected and the excitation spectrum is gapless.
In the orbital phase one has the six-fold anisotropy associated
with the equivalent choices for differently populating orbital
levels in cubic symmetry, as is discussed in the text.}
\label{PHASE}
\end{figure}

\noindent
which gives rise to the phase diagram shown in Fig.~\ref{PHASE}. 
This phase diagram is not quite the same as that
found in Ref. \onlinecite{IHM} for zero temperature. When we have
spin ordering, we may analyze the fourth-order terms, as is done
in Eq. (\ref{QUARTIC}).  That analysis shows that unless the
minimum eigenvector has components of equal magnitude, the
anisotropy favors spin ordering along a $(1,1,1)$ direction. The
condition that the eigenvector be $(-1,1,1)$ is that $y+z=0$. This
can only happen when $\eta=0$.
Then we have isotropy and the mean-field state exhibits rotational
degeneracy. Otherwise, when $\eta \not=0$, the fourth-order terms
give rise to an anisotropy that orients the staggered spin along a
$(1,1,1)$ direction. We should also remind the reader that
fluctuations favor the spin-only state, so that the phase boundary
shown in Fig. \ref{PHASE} will be shifted by fluctuations to
larger positive $\eta$.  In the regime of orbital ordering, we
indicate in Appendix \ref{F} the existence of a six-fold
anisotropy in the variables $a_1({\bf Q})$ and $a_2({\bf Q})$,
such that the six equivalent minima correspond to the six possible
states which are obtained by choosing $N_\alpha=1/3$ for one
coordinate $\alpha$, and  then occupying the two other
orbitals with probability $1/3 \pm \Delta$.

\section{DISCUSSION AND SUMMARY}

The cubic KK model has some very unusual and interesting symmetries
which cause mean-field theory to have some unusual features.
In particular, for the simplest KK Hamiltonian, we found that
mean-field theory leads to criticality for the wavevector-dependent
spin susceptibility associated with orbital $\alpha$ which
is dispersionless along the $q_\alpha$ direction of wavevector.
This result is consistent with the previous observation\cite{PRL}
that the Hamiltonian is invariant against an arbitrary rotation of
the total spin in the orbital $\alpha$ summed over all spins in
any single plane perpendicular to the $\alpha$ axis. This `soft
mode' behavior prevents the development of long-range spin order at
any nonzero temperature,\cite{PRL} {\it even though the system is
a three dimensional one.}

Any perturbation which destroys this
peculiar symmetry will enable the system to develop long-range
spin order.  In particular, we investigate the role of a) spin-orbit
interactions, b) second-neighbor hopping, and c) Hund's rule coupling
in stabilizing long-range spin order.  In the presence of
spin-orbit interaction we find wavevector selection
(because now the spin of different orbitals can not be freely
rotated relative to one another) into a two-sublattice
antiferromagnetic state with a greatly reduced spin magnitude.
Since experiment shows such a reduction,\cite{LTO} this mechanism
may be operative to some extent.  However, as noted previously,\cite{PRL}
the excitation spectrum does not have a gap until further
perturbations are also included.  The mean-field solution is
consistent with this conclusion, because the mean-field state which
minimizes the trial free energy is degenerate with respect to a
global rotation of the staggered spin.

The ordered state which results
when nnn hopping is added to the bare KK Hamiltonian is quite unusual.
In this state, each orbital flavor has a staggered spin moment, but
these three staggered spin moments form a 120$^{\rm o}$ degree state
such that the total staggered spin moment (summed over the three
orbital states) is zero!  It is not immediately obvious how such
long-range order would be observed. Finally, we show that when the
bare KK Hamiltonian is perturbed by the addition of only Hund's rule
coupling, the resulting ordered state may exhibit long-range
{\it antiferromagnetic orbital} order.

One caveat concerning our result should be mentioned.  All our
results are based on a stability analysis of the disordered phase.
If the ordering transition is a discontinuous one, our results
might not reveal such a transition.  In III we will
present results for the temperature-dependence of the various
mean-field solutions. Further analysis of the ordered phase is
needed to obtain a phase diagram at $T=0$, as is done in 
Ref.\onlinecite{IHM}.

It should be emphasized again that all the results in this paper are
based on the assumption that nearest-neighbor bonds along an axis
$\alpha$ are 'inactive', namely that there is no direct hopping
between $\alpha$ orbitals along such bonds. Even within cubic
symmetry, such hopping could still exist, alas with a very small 
hopping energy $t''$. However, as soon as we add such terms, the 
vertical bond in Fig.~\ref{DXY}b becomes active, and Eqs.~(\ref{chid})
and (\ref{chind}) have the additional contributions 
$\Delta \chi^{-1}_{\alpha\alpha} = 2 \epsilon'' c_{\alpha} $ and 
$\Delta \chi^{-1}_{\alpha\beta} = 2 \epsilon''( c_{\alpha}+c_{\beta}) $,
with $\epsilon''=t''^{2}/U$. This introduces dispersion in all
directions, and select order at $\vec{q} = \vec{Q}$. Distortions away
from the cubic structure can enhace $t''$, and stabilize such order 
even further.

One general conclusion from our work is that it is not safe to
associate properties of real experimental systems with properties
of a model Hamiltonian unless one is absolutely sure that the
real system is a realization (at least in all important aspects) of
the model Hamiltonian.  Here the ideal cubic KK Hamiltonian has properties
which are quite different from those observed for systems it supposedly
describes.  What this means is that
it will be necessary to take into account effects that one might have
been tempted to ignore in order to identify a model that is truly
appropriate for experimentally realizable systems.  Alternatively,
perhaps our work will inspire experimentalists to find systems
that are as close as possible to that of the ideal cubic KK
Hamiltonian treated here.  Such systems would have quite
striking and anomalous properties.

\acknowledgments{ ABH thanks NIST for its hospitality during
several visits when this work was done. We acknowledge partial
support from the US-Israel Binational Science Foundation (BSF).
The TAU group is also supported by the German-Israeli Foundation
(GIF).}

\end{multicols}

\appendix

\section{Higher-order terms in the free-energy}
\label{A}

Here we employ Eqs. (\ref{rho}), (\ref{par}), and (\ref{Y}) in
conjunction with Eq. (\ref{XEQ}), to derive general expressions
for the cubic and quartic terms of the free energy.

The `bare' cubic terms in the free-energy arise from Tr[$X^{3}$].
We find
\begin{eqnarray}
&&{\rm
Tr}[X^{3}(i)]=\sum_{\alpha_{i}\beta_{i}}\sum_{\rho_{i}\eta_{i}}{\rm
Tr}\Bigl [ c^{\dagger}_{i\alpha_{1}\rho_{1}}
Y_{\alpha_{1}\rho_{1}\beta_{1}\eta_{1}}(i)c_{i\beta_{1}\eta_{1}}c^{\dagger}_{i\alpha_{2}\rho_{2}}
Y_{\alpha_{2}\rho_{2}\beta_{2}\eta_{2}}(i)c_{i\beta_{2}\eta_{2}}c^{\dagger}_{i\alpha_{3}\rho_{3}}
Y_{\alpha_{3}\rho_{3}\beta_{3}\eta_{3}}(i)c_{i\beta_{3}\eta_{3}}\Bigr ]\nonumber\\
&=&\sum_{\alpha_{i}\rho_{i}}\Bigl
[\Av_{\alpha_{1}\alpha_{2}}(i)\delta_{\rho_{1}\rho_{2}}
+\vec{\Bv}_{\alpha_{1}\alpha_{2}}(i)\cdot\vec{\sigmav}_{\rho_{1}\rho_{2}}\Bigr
]\Bigl [\Av_{\alpha_{2}\alpha_{3}}(i)\delta_{\rho_{2}\rho_{3}}
+\vec{\Bv}_{\alpha_{2}\alpha_{3}}(i)\cdot\vec{\sigmav}_{\rho_{2}\rho_{3}}\Bigr
]\Bigl [\Av_{\alpha_{3}\alpha_{1}}(i)\delta_{\rho_{3}\rho_{1}}
+\vec{\Bv}_{\alpha_{3}\alpha_{1}}(i)\cdot\vec{\sigmav}_{\rho_{3}\rho_{1}}\Bigr
].\label{x3}
\end{eqnarray}
Making use of the identity Eq. (\ref{identity}), this becomes
\begin{eqnarray}
{\rm Tr}[X^{3}(i)]&=&2\sum_{\alpha_{i}}\Bigl \{
\Av_{\alpha_{1}\alpha_{2}}(i)\Av_{\alpha_{2}\alpha_{3}}(i)\Av_{\alpha_{3}\alpha_{1}}(i)
+3\Av_{\alpha_{1}\alpha_{2}}(i)
\vec{\Bv}_{\alpha_{2}\alpha_{3}}(i)\cdot\vec{\Bv}_{\alpha_{3}\alpha_{1}}(i)
\nonumber\\
&&\ \ \ \ \ \ \ \ +i(\vec{\Bv}_{\alpha_{1}\alpha_{2}}(i)
\times\vec{\Bv}_{\alpha_{2}\alpha_{3}}(i))\cdot\vec{\Bv}_{\alpha_{3}\alpha_{1}}(i)\Bigr
\}.\label{bare3}
\end{eqnarray}

The `bare' quartic terms in the free-energy arise from
Tr[$X^{4}$]. We find
\begin{eqnarray}
{\rm
Tr}[X^{4}(i)]&=&\sum_{\alpha_{i}\beta_{i}}\sum_{\rho_{i}\eta_{i}}{\rm
Tr}\Bigl [ c^{\dagger}_{i\alpha_{1}\rho_{1}}
Y_{\alpha_{1}\rho_{1}\beta_{1}\eta_{1}}(i)c_{i\beta_{1}\eta_{1}}
c^{\dagger}_{i\alpha_{2}\rho_{2}}
Y_{\alpha_{2}\rho_{2}\beta_{2}\eta_{2}}(i)c_{i\beta_{2}\eta_{12}}\nonumber\\
&&\times c^{\dagger}_{i\alpha_{3}\rho_{3}}
Y_{\alpha_{3}\rho_{3}\beta_{3}\eta_{3}}(i)c_{i\beta_{3}\eta_{3}}c^{\dagger}_{i\alpha_{4}\rho_{4}}
Y_{\alpha_{4}\rho_{4}\beta_{4}\eta_{4}}(i)c_{i\beta_{4}\eta_{4}}\Bigr ]\nonumber\\
&=&\sum_{\alpha_{i}}\sum_{\rho_{i}}\Big
[\Av_{\alpha_{1}\alpha_{2}}(i)\delta_{\rho_{1}\rho_{2}}
+\vec{\Bv}_{\alpha_{1}\alpha_{2}}(i)\cdot\vec{\sigmav}_{\rho_{1}\rho_{2}}\Bigr
]\Bigl  [\Av_{\alpha_{2}\alpha_{3}}(i)\delta_{\rho_{2}\rho_{3}}
+\vec{\Bv}_{\alpha_{2}\alpha_{3}}(i)\cdot\vec{\sigmav}_{\rho_{2}\rho_{3}}\Bigr
]\nonumber\\
&&\times\Bigl
[\Av_{\alpha_{3}\alpha_{4}}(i)\delta_{\rho_{3}\rho_{4}}
+\vec{\Bv}_{\alpha_{3}\alpha_{4}}(i)\cdot\vec{\sigmav}_{\rho_{3}\rho_{4}}\Bigr
] \Bigl [\Av_{\alpha_{24}\alpha_{1}}(i)\delta_{\rho_{4}\rho_{1}}
+\vec{\Bv}_{\alpha_{4}\alpha_{1}}(i)\cdot\vec{\sigmav}_{\rho_{4}\rho_{1}}\Bigr
].
\end{eqnarray}
Again using the identity Eq. (\ref{identity}), this becomes
\begin{eqnarray}
{\rm Tr}[X^{4}(i)] &=&2\sum_{\alpha_{i}}\Bigl \{\Bigl
(\Av_{\alpha_{1}\alpha_{2}}(i)\Av_{\alpha_{2}\alpha_{3}}(i)
+\vec{\Bv}_{\alpha_{1}\alpha_{2}}(i)\cdot
\vec{\Bv}_{\alpha_{2}\alpha_{3}}(i)\Bigr )\Bigl
(\Av_{\alpha_{3}\alpha_{4}}(i)\Av_{\alpha_{4}\alpha_{1}}(i)
+\vec{\Bv}_{\alpha_{3}\alpha_{4}}(i)\cdot
\vec{\Bv}_{\alpha_{4}\alpha_{1}}(i)\Bigr )\nonumber\\
&+&\Bigl
(\Av_{\alpha_{1}\alpha_{2}}(i)\vec{\Bv}_{\alpha_{2}\alpha_{3}}(i)
+\vec{\Bv}_{\alpha_{1}\alpha_{2}}(i)\Av_{\alpha_{2}\alpha_{3}}(i)
+i\vec{\Bv}_{\alpha_{1}\alpha_{2}}(i)\times\vec{\Bv}_{\alpha_{2}\alpha_{3}}(i)\Bigr
)\nonumber\\
 &&\cdot\Bigl
(\Av_{\alpha_{3}\alpha_{4}}(i)\vec{\Bv}_{\alpha_{4}\alpha_{1}}(i)
+\vec{\Bv}_{\alpha_{3}\alpha_{4}}(i)\Av_{\alpha_{4}\alpha_{1}}(i)
+i
\vec{\Bv}_{\alpha_{3}\alpha_{4}}(i)\times\vec{\Bv}_{\alpha_{4}\alpha_{1}}(i)\Bigr
)\Bigr \}.\label{bare4}
\end{eqnarray}

\section{CUBIC FREE-ENERGY TERMS}
\label{B}

Referring to Eq. (\ref{bare3}), the relevant terms for our purpose
come from  the second and the third terms there. Working in
Fourier space we hence have
\begin{eqnarray}
\delta F&=&-\frac{8\epsilon}{\sqrt{N}}\sum_{{\bf q}_{1}{\bf
q}_{2}}\sum_{\alpha_{1}\alpha_{2}\alpha_{3}}\Bigl
[3\Av_{\alpha_{1}\alpha_{2}}({\bf
q}_{1})\vec{\Bv}_{\alpha_{2}\alpha_{3}}({\bf q}_{2})\cdot
\vec{\Bv}_{\alpha_{3}\alpha_{1}}(-{\bf q}_{1}-{\bf
q}_{2})\nonumber\\
&+&i\vec{\Bv}_{\alpha_{1}\alpha_{2}}({\bf
q}_{1})\times\vec{\Bv}_{\alpha_{2}\alpha_{3}}({\bf q}_{2}) \cdot
\vec{\Bv}_{\alpha_{3}\alpha_{1}}(-{\bf q}_{1}-{\bf q}_{2})\Bigr ].
\end{eqnarray}
When one of the quantities $\Bv$ here acts as the spatially
uniform field [see Eq. (\ref{GEQ})], this expression becomes
\begin{eqnarray}
\delta F&=&-\frac{8\epsilon}{\sqrt{N}}\sum_{{\bf
q}}\sum_{\alpha_{1}\alpha_{2}\alpha_{3}}\Bigl
[3\Av_{\alpha_{1}\alpha_{2}}({\bf
q})\vec{\Bv}_{\alpha_{2}\alpha_{3}}\cdot
\vec{\Bv}_{\alpha_{3}\alpha_{1}}(-{\bf
q})+3\Av_{\alpha_{1}\alpha_{2}}({\bf
q})\vec{\Bv}_{\alpha_{2}\alpha_{3}}(-{\bf q})\cdot
\vec{\Bv}_{\alpha_{3}\alpha_{1}}\nonumber\\
&+&i\vec{\Bv}_{\alpha_{1}\alpha_{2}}\times\vec{\Bv}_{\alpha_{2}\alpha_{3}}({\bf
q}) \cdot \vec{\Bv}_{\alpha_{3}\alpha_{1}}(-{\bf
q})+i\vec{\Bv}_{\alpha_{1}\alpha_{2}}({\bf
q})\times\vec{\Bv}_{\alpha_{2}\alpha_{3}} \cdot
\vec{\Bv}_{\alpha_{3}\alpha_{1}}(-{\bf q})\nonumber\\
&+&i\vec{\Bv}_{\alpha_{1}\alpha_{2}}({\bf q})
\times\vec{\Bv}_{\alpha_{2}\alpha_{3}}(-{\bf q}) \cdot
\vec{\Bv}_{\alpha_{3}\alpha_{1}}\Bigr ],\label{full3}
\end{eqnarray}
where $\vec{\Bv}$ which does not depend on ${\bf q}$ is the
uniform field.

We first consider the terms involving the $\Av$'s. The relevant
contributions come from $\alpha_{3}=\alpha_{1}$ [the first term in
Eq. (\ref{full3}))] and $\alpha_{3}=\alpha_{2}$ [the second term
there]. Hence we find
\begin{eqnarray}
\delta F_A &=&-\frac{24\epsilon}{\sqrt{N}} \sum_{\bf q}
{\sum_{\alpha \beta}}^\prime \Av_{\alpha\beta}({\bf
q})\vec{\Bv}_{\beta\alpha}\cdot\Bigl
(\vec{\Bv}_{\alpha\alpha}(-{\bf q})+\vec{\Bv}_{\beta\beta}(-{\bf
q})\Bigr ),
\end{eqnarray}
where $\sum^{\prime}_{\alpha\beta}$ denotes that
$\alpha\neq\beta$. When we minimize $F_2+\delta F_A$ with respect
to $\Av_{\alpha \beta}({\bf q})$, and use Eqs. (\ref{chind}) and
(\ref{GEQ}), we get the contribution
\begin{eqnarray}
\delta F_A &=& - 72 g_0^2 \epsilon \sum_{\bf q} \sum_{\alpha \beta
\gamma} \epsilon_{\alpha \beta \gamma}^2 [s_{\alpha \gamma}({\bf
q}) + s_{\beta \gamma}({\bf q})]  [s_{\alpha \gamma}(-{\bf q}) +
s_{\beta \gamma}(-{\bf q})] [ 2 + \cos(q_\gamma a) ]^{-1} ,
\end{eqnarray}
where we have defined
\begin{eqnarray}
s_{\alpha \gamma}({\bf q})\equiv \Bv^{\gamma}_{\alpha\alpha}({\bf
q}).\label{sab}
\end{eqnarray}
Also, since we are interested in the free energy to quadratic
order in the order parameters, we have set $kT=2\epsilon/3$.

In this result we want to keep only contributions which
involve the critical variables.  For $s_{\alpha \gamma}({\bf q})$
this means that we sum over ${\bf q}$'s such that
$q_\beta=\pi/a$, for $\beta \not= \alpha$.  Thus for each
$s_{\alpha \gamma}$ the wavevector sum is a sum over the
component $q_\alpha$, with the other components of ${\bf q}$ equal
to $\pi/a$.  We denote this type of sum by $\sum_{q_\alpha}$.
Furthermore for a term involving components $s_{\alpha \gamma}$ and
$s_{\beta \gamma}$ with {\it different} orbitals $\alpha$ and $\beta$,
this sum reduces to
the single wavevector ${\bf Q} = (\pi , \pi, \pi)/a$. So
\begin{eqnarray}
\delta F_A &=& - 144 g_0^2 \epsilon \sum_{\alpha \beta \gamma}
\epsilon_{\alpha \beta \gamma}^2 \Bigl\{ \sum_{q_\alpha}
{s_{\alpha \gamma}({\bf q}) s_{\alpha \gamma}(-{\bf q}) \over 2 +
\cos(q_\gamma a) } + s_{\alpha \gamma}({\bf Q}) s_{\beta
\gamma}({\bf Q}) \Bigr\}  .
\end{eqnarray}
Here we will set $[2 + \cos (q_\gamma a)]=1$ because for $s_{\alpha \gamma}$
(with $\alpha \not= \gamma$) we must have $q_\gamma = \pi/a$.
This term favors ordering at wavevector ${\bf Q}$
with ${\bf s}_\alpha ({\bf Q})$ collinear with $s_{\beta}({\bf Q})$,
where ${\bf s}_\alpha ({\bf Q})$ is a vector with
components $[s_{\alpha x}({\bf Q}) \ ,
s_{\alpha y}({\bf Q}) \ , s_{\alpha z}({\bf Q})]$.

Next we consider the contribution coming from the terms with three
$\Bv$'s in Eq. (\ref{full3}). Here we put one of the ${\bf
q}$-dependent $\Bv$'s to be diagonal in the orbital indices, to
obtain
\begin{eqnarray}
\delta F_{B}&=&-i\frac{36 kT}{\sqrt{N}}\sum_{\bf
q}{\sum_{\alpha\beta}}'\sum_{\alpha_{1}\beta_{1}\gamma}\epsilon_{\alpha_{1}\beta_{1}\gamma}
\Bv_{\alpha\beta}^{\gamma}({\bf
q})(\Bv_{\alpha\alpha}^{\beta_{1}}(-{\bf
q})-\Bv_{\beta\beta}^{\beta_{1}}(-{\bf q}))
\Bv_{\beta\alpha}^{\alpha_{1}}.
\end{eqnarray}
Eliminating  the noncritical $\Bv_{\alpha\beta}^{\gamma}({\bf q})$
variables by minimizing $F_2 + \delta F_B$ with respect to them,
we get
\begin{eqnarray}
\delta F_B &=& - 1296 (g_0kT)^2 {\sum_{\alpha \beta}}^\prime
\sum_{\bf q} \chi _{ \alpha \beta}({\bf q}) [s_{\beta \beta} ({\bf
q}) - s_{\alpha \beta}({\bf q})]  [(s_{\beta \beta} (-{\bf q}) -
s_{\alpha \beta}(-{\bf q})]  ,
\end{eqnarray}
where $\chi$ is given in Eq. (\ref{chind}), and we have used the
definition (\ref{sab}). As before we set $kT=2 \epsilon/3$ and
separate the sums to be only over critical wavevectors for each
orbital spin vector, in which case we have
\begin{eqnarray}
&& \delta F_B = - 144 g_0^2 \epsilon \sum_{\alpha \beta \gamma}
\epsilon_{\alpha \beta \gamma}^2 \Bigl\{ \sum_{q_\alpha}
[s_{\alpha \alpha} ({\bf q}) s_{\alpha \alpha} (-{\bf q}) +
s_{\alpha \beta} ({\bf q}) s_{\alpha \beta} (-{\bf q})] - 2
s_{\alpha \alpha}({\bf Q}) s_{\beta \alpha} ({\bf Q}) \Bigr\} \ .
\end{eqnarray}
Here we noted that $\chi_{\alpha \beta}({\bf q})=\chi_{\alpha
\beta}({\bf Q}) =1/(4 \epsilon)$ because this component of $\chi$
depends on $q_\gamma$ which is always $\pi/a$ in the summation
over wavevector.

In summary the total contribution to the quadratic free energy at
order $\lambda^2$ is
\begin{eqnarray}
 F_2^{(3)} = \delta F_A + \delta F_B &=& - C_0 \sum_{\alpha \beta
\gamma} \epsilon_{\alpha \beta \gamma}^2 \Bigl\{ \Bigl( s_{\alpha
\gamma}({\bf Q})s_{\beta \gamma}({\bf Q})  - 2s_{\beta \beta}
({\bf Q}) s_{\alpha \beta}({\bf Q}) \Bigr)\nonumber\\
& +& \sum_{q_\alpha} \Bigl( s_{\alpha \gamma}({\bf q}) s_{\alpha
\gamma}(-{\bf q}) + s_{\alpha \alpha}({\bf q}) s_{\alpha
\alpha}(-{\bf q}) + s_{\alpha \beta}({\bf q}) s_{\alpha
\beta}(-{\bf q})] \Bigr) \Bigr\}   ,
\end{eqnarray}
where we set $kT=2\epsilon/3$ and $C_0 = 144 g_0^2 \epsilon$.

\section{QUARTIC TERMS IN THE FREE ENERGY}
\label{C}

Now we look at fourth order terms.  These involve two critical
order parameters and two powers of $\lambda$. Therefore, we pick
from Eq. (\ref{bare4}) all terms involving at least two powers of
$\Bv$. Since two of the factors $\Bv$ in each term have to be
$\vec{\Bv}_{\alpha\beta}=-\vec{\Bv}_{\beta\alpha}$, with
$\alpha\neq\beta$, [see Eq. (\ref{GEQ})], we see that the terms
involving a single power of $\Av$ vanish. Thus we have to consider
the expression
\begin{eqnarray}
&&36kT\sum_{i}\sum_{\stackrel{\alpha_{1}\alpha_{2}}{\alpha_{3}\alpha_{4}}}\Bigl
(4\Av_{\alpha_{1}\alpha_{2}}\Av_{\alpha_{2}\alpha_{3}}\vec{\Bv}_{\alpha_{3}\alpha_{4}}\cdot
\vec{\Bv}_{\alpha_{4}\alpha_{1}}+2\Av_{\alpha_{1}\alpha_{2}}\Av_{\alpha_{3}\alpha_{4}}
\vec{\Bv}_{\alpha_{2}\alpha_{3}}\cdot
\vec{\Bv}_{\alpha_{4}\alpha_{1}}\nonumber\\
&+&(\vec{\Bv}_{\alpha_{1}\alpha_{2}}\cdot\vec{\Bv}_{\alpha_{2}\alpha_{3}})(
\vec{\Bv}_{\alpha_{3}\alpha_{4}}\cdot\vec{\Bv}_{\alpha_{4}\alpha_{1}})
-(\vec{\Bv}_{\alpha_{1}\alpha_{2}}\times\vec{\Bv}_{\alpha_{2}\alpha_{3}})\cdot(
\vec{\Bv}_{\alpha_{3}\alpha_{4}}\times\vec{\Bv}_{\alpha_{4}\alpha_{1}})\Bigr
),\label{full4}
\end{eqnarray}
where $\Av$ and $\Bv$ are functions of the site index $i$. The
first two members of Eq. (\ref{full4}) are calculated for the case
in which the $\Av$'s are critical, and the $\Bv$'s are given by
Eq. (\ref{GEQ}). Denoting their contribution to the self-energy by
$\delta F_{2}^{(1)}$, we find
\begin{eqnarray}
\delta F_{2}^{(1)}&=&36kT\sum_{i}\sum_{\alpha\beta}\Bigl
(4\Av_{\alpha\alpha}^{2}(i)+2\Av_{\alpha\alpha}(i)\Av_{\beta\beta}(i)\Bigr
)\vec{\Bv}_{\alpha\beta}\cdot\vec{\Bv}_{\beta\alpha}\nonumber\\
&=&36kTg_{0}^{2}\sum_{\alpha\beta\gamma}\epsilon^{2}_{\alpha\beta\gamma}
\Bigl
((4\Av_{\alpha\alpha}^{2}(i)+2\Av_{\alpha\alpha}(i)\Av_{\beta\beta}(i)\Bigr
)=
216kTg_{0}^{2}\sum_{i}(a_{1}^{2}(i)+a_{2}^{2}(i)),\label{QUADORB}
\end{eqnarray}
where in the last step we have used Eq. (\ref{map}).

The contribution of the remaining two members of Eq. (\ref{full4})
is denoted $\delta F_{2}^{(2)}$. Here we have to take two of the
$\Bv$'s as critical, while the other two are given by Eq.
(\ref{GEQ}). To shorten notations, we denote here the critical
$\Bv$ as $\Bv(i)$, while the non-critical one is simply written as
$\Bv$. We have
\begin{eqnarray}
&&\delta F_{2}^{(2)}=72kT\sum_{i}\sum_{\alpha\beta}\Bigr
[(\vec{\Bv}_{\alpha\alpha}(i)\cdot \vec{\Bv}_{\alpha\alpha}(i))
(\vec{\Bv}_{\alpha\beta}\cdot \vec{\Bv}_{\beta\alpha})+
(\vec{\Bv}_{\alpha\alpha}(i)\cdot \vec{\Bv}_{\alpha\beta})
(\vec{\Bv}_{\alpha\alpha}(i)\cdot \vec{\Bv}_{\beta\alpha})\nonumber\\
&+& (\vec{\Bv}_{\alpha\alpha}(i)\cdot \vec{\Bv}_{\alpha\beta})
(\vec{\Bv}_{\beta\beta}(i)\cdot
\vec{\Bv}_{\beta\alpha})-(\vec{\Bv}_{\alpha\alpha}(i)\times\vec{\Bv}_{\alpha\beta})
\cdot (\vec{\Bv}_{\beta\beta}(i)\times\vec{\Bv}_{\beta\alpha})
+(\vec{\Bv}_{\alpha\alpha}(i)\times\vec{\Bv}_{\alpha\beta}) \cdot
(\vec{\Bv}_{\alpha\alpha}(i)\times\vec{\Bv}_{\beta\alpha})\Bigr ].
\end{eqnarray}
Making again use of Eq. (\ref{GEQ}), this expression becomes
\begin{eqnarray}
\delta
F_{2}^{(2)}&=&72kTg_{0}^{2}\sum_{i}\sum_{\alpha\beta\gamma}\epsilon^{2}_{\alpha\beta\gamma}
\Bigl
[2\sum_{\nu}\Bv_{\alpha\alpha}^{\nu}(i)\Bv_{\alpha\alpha}^{\nu}(i)-
\sum_{\nu}\Bv_{\alpha\alpha}^{\nu}(i)\Bv_{\beta\beta}^{\nu}(i) +2
\Bv_{\alpha\alpha}^{\gamma}(i)\Bv_{\beta\beta}^{\gamma}(i)\Bigr ].
\end{eqnarray}
Transforming to Fourier space, noting that only  the first term
here contains ${\bf q}$ while in the other two we must necessarily
have ${\bf q}={\bf Q}$, (because they involve simultaneous
criticality of two flavors), we obtain
\begin{eqnarray}
\delta F_{2}^{(2)}&=&72kTg_{0}^{2}\Bigl [2\sum_{\bf
q}\sum_{\alpha\beta}s_{\alpha\beta}({\bf q})s_{\alpha\beta}(-{\bf
q})+\sum_{\alpha\beta\gamma}\epsilon^{2}_{\alpha\beta\gamma}\Bigl
(2s_{\alpha\gamma}({\bf Q})s_{\beta\gamma}({\bf
Q})-\sum_{\nu}s_{\alpha\nu}({\bf Q})s_{\beta\nu}({\bf Q})\Bigr
)\Bigr ],
\end{eqnarray}
where we have used the definition Eq. (\ref{sab}). The total
contribution to the free energy from quartic terms is then
\begin{eqnarray}
F_{2}^{(4)}=\delta F_{2}^{(1)}+\delta F_{2}^{(2)}.
\end{eqnarray}

\section{MEAN-FIELD THEORY FOR nnn HOPPING}
\label{D}

Starting from Eq. (\ref{VKK}), we may write the perturbation due
to next-nearest-neighbors in the form
\begin{eqnarray}
V_{KK}&=&\epsilon
'\sum_{ij}\sum_{\alpha\beta\delta}\gamma_{\alpha}(i,j)\sum_{\rho
\eta}
 \epsilon^{2}_{\alpha\beta\delta}\Bigl
[c^{\dagger}_{i\beta\rho}c_{i\delta\eta}c^{\dagger}_{j\beta\eta}c_{j\delta\rho}
+c^{\dagger}_{i\beta\rho}c_{i\beta\eta}c^{\dagger}_{j\delta\eta}c_{j\delta\rho}\Bigr
].
\end{eqnarray}
Within our mean-field theory, the averages are taken separately on
the operators belonging to the site $i$, and those belonging to
site $j$. The required averages are then given in Eq.
(\ref{AVERAGE}). The following contribution to the trial energy
$U$ is then
\begin{eqnarray}
\langle V_{KK}\rangle &=&2\epsilon
'\sum_{ij}\sum_{\alpha\beta\delta}\gamma_{\alpha}(i,j)
\epsilon^{2}_{\alpha\beta\delta}\Bigl
[\Av_{\delta\beta}(i)\Av_{\delta\beta}(j)+\vec{\Bv}_{\delta\beta}(i)\cdot\vec{\Bv}_{\delta\beta}(j)
+\Av_{\beta\beta}(i)\Av_{\delta\delta}(j)+\vec{\Bv}_{\beta\beta}(i)\cdot\vec{\Bv}_{\delta\delta}(j)
\Bigr ].
\end{eqnarray}
Transforming to Fourier space, noting that each site has four
next-nearest neighbors in each $\alpha$-plane, we obtain
\begin{eqnarray}
\langle V_{KK}\rangle &=&8\epsilon '\sum_{\bf q
}\sum_{\alpha\beta\delta}\epsilon^{2}_{\alpha\beta\delta}c_{\beta}c_{\delta}\Bigl
[\Av_{\delta\beta}({\bf q} )\Av_{\delta\beta}(-{\bf q}
)+\vec{\Bv}_{\delta\beta}({\bf
q})\cdot\vec{\Bv}_{\delta\beta}(-{\bf q}) +\Av_{\beta\beta}({\bf
q} )\Av_{\delta\delta}(-{\bf q} )+\vec{\Bv}_{\beta\beta}({\bf q}
)\cdot\vec{\Bv}_{\delta\delta}(-{\bf q}) \Bigr ],\label{VKKAVER}
\end{eqnarray}
where $c_{\beta}=\cos (q_{\beta}q)$. The result Eq.
(\ref{VKKAVER}) is now added to Eq. (\ref{FEQ}), in order to
obtain the modifications in the susceptibility tensor. Specifying
to the diagonal order-parameters $\Av_{\alpha\alpha}$ and
$\vec{\Bv}_{\alpha\alpha}$, the susceptibility tensor becomes [see
Eq. (\ref{chid})]
\begin{eqnarray}
 \chiv ({\bf q})^{-1}  &=& \left[ \begin{array}
{c c c} 12 kT + 4 \epsilon (c_y + c_z) & 8 \epsilon^\prime c_x c_y
&
8 \epsilon^\prime c_x c_z \\
8 \epsilon^\prime c_x c_y & 12kT + 4 \epsilon(c_x+c_z)
& 8 \epsilon^\prime c_y c_z \\
8 \epsilon^\prime c_x c_z & 8 \epsilon^\prime c_y c_z
& 12kT + 4 \epsilon(c_x+c_y) \\
\end{array} \right].
\end{eqnarray}

Now we look at the most critical wavevector, which here is
${\bf Q}$.  There we have
\begin{eqnarray}
\chiv ({\bf Q})^{-1} &=& \left[ \begin{array} {c c c}
12 kT - 8 \epsilon  & 8 \epsilon^\prime & 8 \epsilon^\prime \\
8 \epsilon^\prime & 12kT - 8 \epsilon & 8 \epsilon^\prime  \\
8 \epsilon^\prime & 8 \epsilon^\prime & 12kT - 8 \epsilon \\
\end{array} \right]  .\label{spinmatrix}
\end{eqnarray}
We begin with the analysis of the susceptibility tensor of the
spin order parameters, which are given by the elements of
$\Bv_{\alpha\alpha}$. Then we can use the matrix
(\ref{spinmatrix}). The minimum eigenvalue is
\begin{eqnarray}
\lambda &=& 12 kT - 8 \epsilon - 8 \epsilon^\prime
,\label{minlam}
\end{eqnarray}
which gives
\begin{eqnarray}
kT_c &=&  2 \epsilon/3 + 2\epsilon^\prime/3  .
\end{eqnarray}
Correspondingly, there are two degenerate eigenvectors:
\begin{eqnarray}
|1 \rangle &=& (0, 1 , -1)/\sqrt 2 \ , \ \  |2 \rangle = (2, -1 ,
-1)/\sqrt 6 \  .
\end{eqnarray}
To avoid confusion between orbital and spin labels, we will here
denote the orbital states $x$,  $y$, and $z$ by $a$, $b$, and $c$.
Then in terms of normal mode vector $\xiv$ and $\rhov$ we have the
orbital spin vectors as
\begin{eqnarray}
{\bf s}_{a}=-\frac{2}{\sqrt{6}}\rhov\ ,\ \ {\bf
s}_{b}=\frac{1}{\sqrt{6}}\rhov +\frac{1}{\sqrt{2}}\xiv\ ,\ \ {\bf
s}_{c}=\frac{1}{\sqrt{6}}\rhov -\frac{1}{\sqrt{2}}\xiv\ ,
\end{eqnarray}
with
\begin{eqnarray}
{\bf s}_a^2 &=& \frac{2}{3} \rhov^2 \ , \ \ {\bf s}_b^2 ={1 \over
2} \xiv^2 + {1 \over 6} \rhov^2 + \frac{1}{\sqrt{3}}\xiv \cdot
\rhov  \ ,\ \ {\bf s}_c^2 = {1 \over 2} \xiv^2 + {1 \over 6}
\rhov^2 - \frac{1}{\sqrt{3}}\xiv \cdot \rhov   \ .
\end{eqnarray}
Evaluating the fourth-order free energy [see Eq. (\ref{FFOUR})]
relevant to the spin-order parameters, we find
\begin{eqnarray}
\Bigl (\sum_{\mu}{\bf s}_{\mu}^{2}\Bigr )^{2}-\sum_{\mu<\nu}{\bf
s}_{\mu}^{2}{\bf
s}_{\nu}^{2}=\frac{3}{4}(\xiv^{2}+\rhov^{2})^{2}-\frac{1}{3}(\xiv\times\rhov
)^{2}.
\end{eqnarray}
What we see is that the fourth-order term does not select a
particular direction for order.  We have three angles which
describe the degenerate manifold.  For a given value of $\xiv^2 +
\rhov^2$, we optimize the term $(\xiv \times \rhov )^2$  by taking
$| \xiv| = |\rhov |$ and making $\xiv$ perpendicular to $\rhov$.
So, it takes two angles to specify $\xiv$ (given that its length
is fixed) and then we have one angle to specify $\rhov$, given
that $|\rhov| = |\xiv|$ and it is perpendicular to $\xiv$. We now
discuss what this choice of order parameters means for the spin
vectors. First note that
\begin{eqnarray}
{\bf s}_\alpha^2 &=& {\bf s}_\beta^2 = {\bf s}_\gamma^2 =
2 \xiv^2 /3 \ .
\end{eqnarray}
Also we see that the three orbital spin vectors obey
\begin{eqnarray}
{\bf s}_a \cdot {\bf s}_b &=&
{\bf s}_a \cdot {\bf s}_c =
{\bf s}_b \cdot {\bf s}_c = - \xiv^2/3 \ .
\end{eqnarray}
The three vectors each make a 120$^{\rm o}$ angle with each other
and must therefore lie in a single plane.
We can fix, say, ${\bf s}_\alpha$.  This accounts for two
angles.  Then the other two spin vectors require another
angle to tell which plane they lie in.  Note that there is
zero net staggered moment.  There is long-range spin order,
but not of any simple type.

Next we analyze the susceptibility tensor of the occupation order
parameters, which are given by the elements of
$\Av_{\alpha\alpha}$. Since the matrix $\Av_{\alpha\alpha}$ is
traceless, we use the parametrization Eq. (\ref{map}) to obtain
from Eq. (\ref{spinmatrix}) the 2$\times$2 matrix
\begin{eqnarray}
 \chi_{\mu \nu} ({\bf q})^{-1} =  \left[
\begin{array}
{cc}12kT+\frac{2\epsilon}{3}(5c_{x}+5c_{y}+2c_{z})+\frac{8\epsilon
'}{3}(c_{x}c_{y}-2c_{y}c_{z}-2c_{z}c_{x})&\frac{2\epsilon}{\sqrt{3}}(c_{y}-c_{x})+\frac{8\epsilon
'}{\sqrt{3}}c_{z}(c_{y}-c_{x})\\
\frac{2\epsilon}{\sqrt{3}}(c_{y}-c_{x})+\frac{8\epsilon
'}{\sqrt{3}}c_{z}(c_{y}-c_{x})&12kT+2\epsilon
(c_{x}+c_{y}+2c_{z})-8\epsilon 'c_{x}c_{y}\end{array}\right ].
\end{eqnarray}
This gives a minimum eigenvalue identical to that of Eq.
(\ref{minlam}),
which yields the same instability temperature as for the spin-only
states. However, in the absence of second-neighbor coupling, the
spin-only states are favored by fluctuations, \cite{IV} so that
choice should be maintained for infinitesimal next-nearest
neighbor hopping.  (The situation could change when the
next-nearest neighbor hopping exceed some threshold value.)

\section{Derivation of the Hund's rule Hamiltonian}

\label{E}

The Coulomb exchange terms for the $t_{2g}$-states can be written
in the form \cite{IHM}
\begin{eqnarray}
{\cal
H}_{cex}=\frac{J_{H}}{2}\sum_{i}\sum_{\stackrel{\alpha\beta}{\alpha\neq\beta}}
\sum_{\sigma\sigma '}\Bigl
(c_{i\alpha\sigma}^{\dagger}c_{i\alpha\sigma '
}^{\dagger}c_{i\beta\sigma
'}c_{i\beta\sigma}+c_{i\alpha\sigma}^{\dagger}c_{i\beta\sigma '
}^{\dagger}c_{i\alpha\sigma
'}c_{i\beta\sigma}-2c_{i\alpha\sigma}^{\dagger}c_{i\beta\sigma '
}^{\dagger}c_{i\beta\sigma '}c_{i\alpha\sigma}\Bigr ),
\end{eqnarray}
where $J_{H}$ is the Hund's rule coupling. Adding ${\cal H}_{cex}$
to the Hamiltonian Eq. (\ref{HHUB}), the perturbation expansion in
power of the transfer integrals $t$ now contains a term of the
order $t^{2}J_{H}/U^{2}$, which reads
\begin{eqnarray}
\delta {\cal H}_{KK}&=&\frac{t^{2}J_{H}}{U^{2}}\sum_{\langle
ij\rangle}\sum_{\beta\gamma\neq\langle
ij\rangle}\sum_{\sigma\sigma '}\Bigl
(c^{\dagger}_{i\gamma\sigma}c_{i\beta\sigma}c^{\dagger}_{j\gamma\sigma
'}c_{j\beta\sigma '}-c^{\dagger}_{i\gamma\sigma '
}c_{i\beta\sigma}c^{\dagger}_{j\gamma\sigma }c_{j\beta\sigma '}+
c^{\dagger}_{i\gamma\sigma}c_{i\beta\sigma}c^{\dagger}_{j\beta\sigma
'}c_{j\gamma\sigma '}\nonumber\\
&-& c^{\dagger}_{i\beta\sigma '
}c_{i\beta\sigma}c^{\dagger}_{j\gamma\sigma }c_{j\gamma\sigma '}
-2c^{\dagger}_{i\beta\sigma}c_{i\beta\sigma}c^{\dagger}_{j\gamma\sigma
'}c_{j\gamma\sigma '}+2c^{\dagger}_{i\gamma\sigma '
}c_{i\beta\sigma}c^{\dagger}_{j\beta\sigma }c_{j\gamma\sigma
'}\Bigr ).
\end{eqnarray}
Taking the thermal averages using Eq. (\ref{AVERAGE}) we find
\begin{eqnarray}
\langle\delta {\cal H}_{KK}\rangle
&=&\frac{t^{2}J_{H}}{U^{2}}\sum_{\langle
ij\rangle}\sum_{\beta\gamma\neq\langle ij\rangle}\Bigl
(2\Av_{\beta\gamma}(i)\Av_{\beta\gamma}(j)+8\Av_{\beta\gamma}(i)\Av_{\gamma\beta}(j)-10\Av_{\beta\beta}(i)
\Av_{\gamma\gamma}(j)\nonumber\\
&-&2\vec{\Bv}_{\beta\gamma}(i)\cdot\vec{\Bv}_{\beta\gamma}(j)
+4\vec{\Bv}_{\beta\gamma}(i)\cdot\vec{\Bv}_{\gamma\beta}(j)-
2\vec{\Bv}_{\beta\beta}(i)\cdot\vec{\Bv}_{\gamma\gamma}(j)\Bigr ).
\end{eqnarray}
where terms independent of the order-parameters were omitted.

\section{SIXTH-ORDER ANISOTROPY in the orbital sector}

\label{F}

At fourth-order, the terms in $a_1(i)$ and $a_2(i)$ are
proportional to $[a^{2}_1(i)+a_2^{2}(i)]^2$ [see Eq.
({\ref{FFOUR})],
and there is complete isotropy in $a_1-a_2$ space. However, this
isotropy must be broken in view of the special role played by the
directions along the cubic crystal axes. This symmetry is found in
the sixth-order terms, as we now show. There are several
contributions to the free energy at sixth order in $a_{1}(i)$ and
$a_{2}(i)$, some of which involve coupling to non-critical
variables. To illustrate the symmetry of these terms we explicitly
consider only the ``direct" terms arising from Eq. (\ref{XEQ}),
from which we have
\begin{eqnarray}
\delta F &=& a \sum_i {\rm Tr} X^{6}(i) \ ,
\end{eqnarray}
where $a$ is a numerical coefficient times $kT$.
Thus we write
\begin{eqnarray}
\delta F = \sum_i {\rm Tr} \Biggl[ \sum_{\alpha \beta \rho \eta}
c_{i \alpha \rho}^\dagger A_{\alpha \beta}(i) \delta_{\rho, \eta}
c_{i \beta \eta} \Biggr]^6 = a \sum_i {\rm tr} {\bf A}^6 (i) \ ,
\end{eqnarray}
where here the trace operation, indicated by ``tr,'' refers to a
diagonal sum over the indices of the matrix ${\bf A}$, as
contrasted to the trace used elsewhere in this paper over the 6
$t_{2g}$-states.  Using Eq. (\ref{map}), this yields
\begin{eqnarray}
\delta F &=& a \sum_i \Biggl[ \left( {a_1(i)+\sqrt 3 a_2(i) \over
\sqrt 6 } \right)^6  + \left( {a_1(i)-\sqrt 3 a_2(i) \over \sqrt 6
} \right)^6 + \left( {-2 a_1(i) \over \sqrt 6} \right)^6 \Biggr] \
.
\end{eqnarray}
Now, since we are only interested in how this term affects the
critical variables, we may replace $\sqrt N a_n(i)$ by $a_n({\bf
Q})$, which we denote $a_n$.  Then we may write
\begin{eqnarray}
\delta F &=& {a \over 36N^2} \Biggl[ 10 [a_1^2 + a_2^2]^3 + a_1^6
- 15 a_1^4 a_2^2 + 15 a_1^2a_2^4 -a_2^6 \Biggr] \ .
\end{eqnarray}
To clarify the anisotropy of this form we set $a_1=r \cos
\theta_{\bf Q}$ and $a_2=r \sin \theta_{\bf Q}$, in which case
\begin{eqnarray}
\delta F = {ar^6 \over 36 N^2} [ 10 + \cos (6 \theta_{\bf Q}) ] \
.\label{f6}
\end{eqnarray}
This free energy has minima at the angles $\theta_{\bf Q}= \pi/2 +
n \pi /3$, for $n=0,1 \dots 5$.  These correspond to $a_1= -r \sin
(n \pi /3)$ and $a_2 = r \cos(n \pi /3)$. For $n=0$, only $a_2$ is
nonzero.  From Eqs. (\ref{NEQ}) one sees that this corresponds to
$\langle N_z(i) \rangle =1/3$,  and having $N_x(i)-N_y(i)$
oscillate at wavevector ${\bf Q}$ with an amplitude proportional
to $r$.  By similarly analyzing the other minima, one concludes
that these six minima correspond to the six ways one can chose
indices so that $\langle N_\alpha (i)\rangle =1/3$ and $\langle
N_\beta (i) - N_\gamma (i) \rangle$ oscillate at wavevector ${\bf
Q}$.  (There are three ways to choose $\alpha$ and two ways to fix
the phase of the orbital density wave.) However,  additional
contributions to the free energy might make the coefficient of the
cosine term in Eq. (\ref{f6}) negative, in which case the minima
occur for $\theta_{\bf Q}=n\pi /3$. Now for $n=0$ only $a_{1}$ is
nonzero, and, from Eqs. (\ref{NEQ}), this corresponds to
$N_x(i)=N_y(i)=\frac{1}{3}+\delta (i)$, and
$N_{z}(i)=\frac{1}{3}-2\delta (i)$, where $\delta (i)$ oscillates
at wavevector ${\bf Q}$. The other minima correspond to cyclic
permutations of  coordinate axes consistent with cubic symmetry.

\begin{multicols}{2}

\end{multicols}{2}

\begin{thebibliography} {99}
\bibitem{rem}
Note that this result is different from the one reported in Ref.
\onlinecite{HUND}.
\end{thebibliography}
\end{document}